%% file: main_mm_integral.tex
\documentclass[final,3p,times,twocolumn,authoryear]{elsarticle}

\usepackage{amssymb}

\usepackage{xspace}
\usepackage{aas_macros}
\usepackage{xcolor}
\usepackage{hyperref}
\usepackage{rotating}
\usepackage{longtable}

\usepackage{array}
\newcolumntype{L}[1]{>{\raggedright\let\newline\\\arraybackslash\hspace{0pt}}m{#1}}
\newcolumntype{C}[1]{>{\centering\let\newline\\\arraybackslash\hspace{0pt}}m{#1}}
\newcolumntype{R}[1]{>{\raggedleft\let\newline\\\arraybackslash\hspace{0pt}}m{#1}}
\graphicspath{{images/}}

\journal{New Astronomy}

\newcommand{\intg}{INTEGRAL\xspace}

\begin{document}

\begin{frontmatter}



\title{Multi-messenger astronomy with INTEGRAL}

\author[isdc]{Carlo Ferrigno}
\author[isdc]{Volodymyr Savchenko}
\author[cnrs]{Alexis Coleiro}
\author[iap]{Francesca Panessa}
\author[iap]{Angela Bazzano}
\author[isdc]{Enrico Bozzo}
\author[dtu]{J\'er\^ome Chenevez}
\author[cab]{Albert Domingo}
\author[dub]{Maeve Doyle}
\author[cnrs,irfu]{Andrea Goldwurm}
\author[cea]{Diego G\"otz}
\author[irap]{Elisabeth Jourdain}
\author[mpe]{A.~von Kienlin}
\author[estec]{Erik Kuulkers}
\author[iasfmi]{Sandro Mereghetti}
\author[dub]{Antonio Martin-Carrillo}
\author[iap]{Lorenzo Natalucci}
\author[iap]{Francesca Onori}
\author[iap]{James Rodi}
\author[irap]{Jean-Pierre Roques}
\author[esac]{Celia S\'anchez-Fern\'andez}
\author[iap]{Pietro Ubertini}

\address[isdc]{ISDC, Department of astronomy, University of Geneva,
                chemin d'\'Ecogia, 16; 1290 Versoix, Switzerland}
\address[cnrs]{Universit\'e de Paris, CNRS, AstroParticule et Cosmologie, F-75013, Paris}
\address[iap]{INAF -- Institute for Space Astrophysics and Planetology, Via Fosso del Cavaliere 100, I-00133 Rome, Italy}
\address[dtu]{DTU Space, National Space Institute Elektrovej - Building 327 DK-2800 Kongens Lyngby Denmark}
\address[cab]{Centro de Astrobiolog{\'{\i}}a -- Departamento de Astrof{\'{\i}}sica (CSIC-INTA), 28692 Villanueva de la Ca{\~n}ada, Madrid, Spain}
\address[dub]{School of Physics, University College Dublin, Belfield, Dublin 4, Ireland}
\address[irfu]{IRFU, CEA, Universit\'e Paris-Saclay, F-91191 Gif-sur-Yvette, France}
\address[cea]{AIM-CEA/DRF/Irfu/D\'epartement d'Astrophysique, CNRS, Universit\'e Paris-Saclay, Universit\'e de Paris, \\
	\hspace{0.05cm} Orme des Merisiers, F-91191 Gif-sur-Yvette, France}
\address[irap]{CNRS; IRAP; 9 Av. colonel Roche, BP 44346, F-31028 Toulouse cedex 4, France\\
	\hspace{0.05cm} Universit\'e de Toulouse; UPS-OMP; IRAP;  Toulouse, France\\}
\address[iasfmi]{INAF -- Istituto di Astrofisica Spaziale e Fisica Cosmica, Via A. Corti 12, I-20133 Milano, Italy}
\address[mpe]{Max-Planck-Institut f\"{u}r extraterrestrische Physik, Giessenbachstrasse 1, D-85748 Garching, Germany}
\address[estec]{ESA/ESTEC, Keplerlaan 1, 2201 AZ Noordwijk, The Netherlands}
\address[esac]{European Space Astronomy Centre (ESA/ESAC), Science Operations Department,\\
	\hspace{0.05cm}P.O. Box 78, E-28691, Villanueva de la Ca\~{n}ada, Madrid, Spain}

\begin{abstract}

At the time of defining the science objectives of the INTernational Gamma-Ray Astrophysics Laboratory (INTEGRAL),
such a rapid and spectacular development of multi-messenger astronomy could not have been predicted,
with new impulsive phenomena becoming accessible through different channels.
Neutrino telescopes have routinely detected energetic neutrino events coming
from unknown cosmic sources since 2013.
Gravitational wave detectors opened a novel window on the sky in 2015 with the detection of the merging
of two black holes and in 2017 with the merging of two neutron stars, followed by signals in the full
electromagnetic range.
Finally, since 2007, radio telescopes detected extremely intense and short burst of radio waves, known as Fast Radio Bursts (FRBs)
whose origin is for most cases extragalactic, but enigmatic.
The exceptionally robust and
versatile design of the INTEGRAL mission has allowed researchers to exploit data collected not only with the
pointed instruments, but also with the active cosmic-ray shields
of the main instruments to detect impulses of gamma-rays in
coincidence with unpredictable phenomena. The full-sky coverage, mostly unocculted by the Earth,
the large effective area, the stable background, and the high duty cycle (85\%) put INTEGRAL in a
privileged position to give a major contribution to multi-messenger astronomy.
In this review, we describe how INTEGRAL has provided upper limits
on the gamma-ray emission from black-hole binary mergers,
detected a short gamma-ray burst in coincidence with a binary neutron star merger, contributed to define the spectral
energy distribution of a blazar associated with a neutrino event,
set upper limits on impulsive and steady gamma-ray emission from cosmological FRBs,
and detected a magnetar flare associated with fast radio bursting emission.

\end{abstract}

\begin{keyword}

fast radio bursts \sep gravitational waves \sep neutrinos \sep  stars: neutron



\end{keyword}

\end{frontmatter}

\tableofcontents

\section{Introduction}
\label{Sec:Intro}

Multi-messenger astronomy was a niche concept at the time of conception and launch of the INTernational Gamma-Ray
Astrophysics Laboratory (\intg). As a consequence, this review is based on relatively recent developments. The first reports of high-energy neutrinos date back to 2013 \citep{neutrino1,neutrino2} and the living alerts started in 2016 \citep{Aartsen2017} allowing rapid follow-up in search of serendipitous signals in gamma-rays.
The first detection of gravitational waves was achieved in September 2015 \citep{Abbott2016c} and opened a new window on the energetic universe. Fast radio bursts are other impulsive events first reported in 2007 \citep{Lorimer2007}, proposed to be associated with magnetar flares some years later \citep[e.g.,][]{Popov2013}, and with a confirmed association found only in 2020 \citep{Mereghetti2020}.

\intg has an 85\% duty cycle, linked to the 2.7 d elliptical orbit, during which the satellite passes through the Earth's radiation belts and instruments are switched off for safety.\intg is well equipped to search for serendipitous events in gamma-rays thanks to the all-sky
coverage of some of its instruments and is designed to follow up unknown events with its large field of view and good sensitivity in hard X-rays and gamma-rays (see Sect.~\ref{Sec:INST} for further details).
As part of the core science for \intg, the \intg Science Data Centre (ISDC) routinely performs a search for transient phenomena in the X- and gamma-ray bands. In particular,
the IBAS system detects a gamma-ray burst (GRB) in the IBIS field
of view every second month on average\footnote{This rate is variable throughout the mission lifetime, as can be seen from the \href{http://ibas.iasf-milano.inaf.it/}{IBAS online catalog}.} \citep{Mereghetti2003,Mereghetti2004}, and once per week in the
anti-coincidence shield of the spectrometer (SPI-ACS)\footnote{A full catalog of SPI-ACS bursts is available at \href{https://www.isdc.unige.ch/integral/science/grb\#ACS}{the SPI-ACS online catalog}.}.
When in the field of view, GRBs can be localized
with a precision of a few arcminutes, whereas no localization is provided for SPI-ACS bursts. However, by combining the information from the different \intg detectors, it is possible to have a rough constraint on the sky region from which a signal is emitted \citep{Savchenko2017}. The SPI-ACS signal is routinely used in combination with other detectors for triangulation measurements to localize
GRBs through the Interplanetary Network \citep[e.g][]{Hurley2013}; this is particularly relevant for events not detected by Swift-BAT or INTEGRAL-IBIS, or for which Fermi-GBM localization is
not possible.

\begin{figure*}
	\label{fig:combined}
	\includegraphics[width=\textwidth]{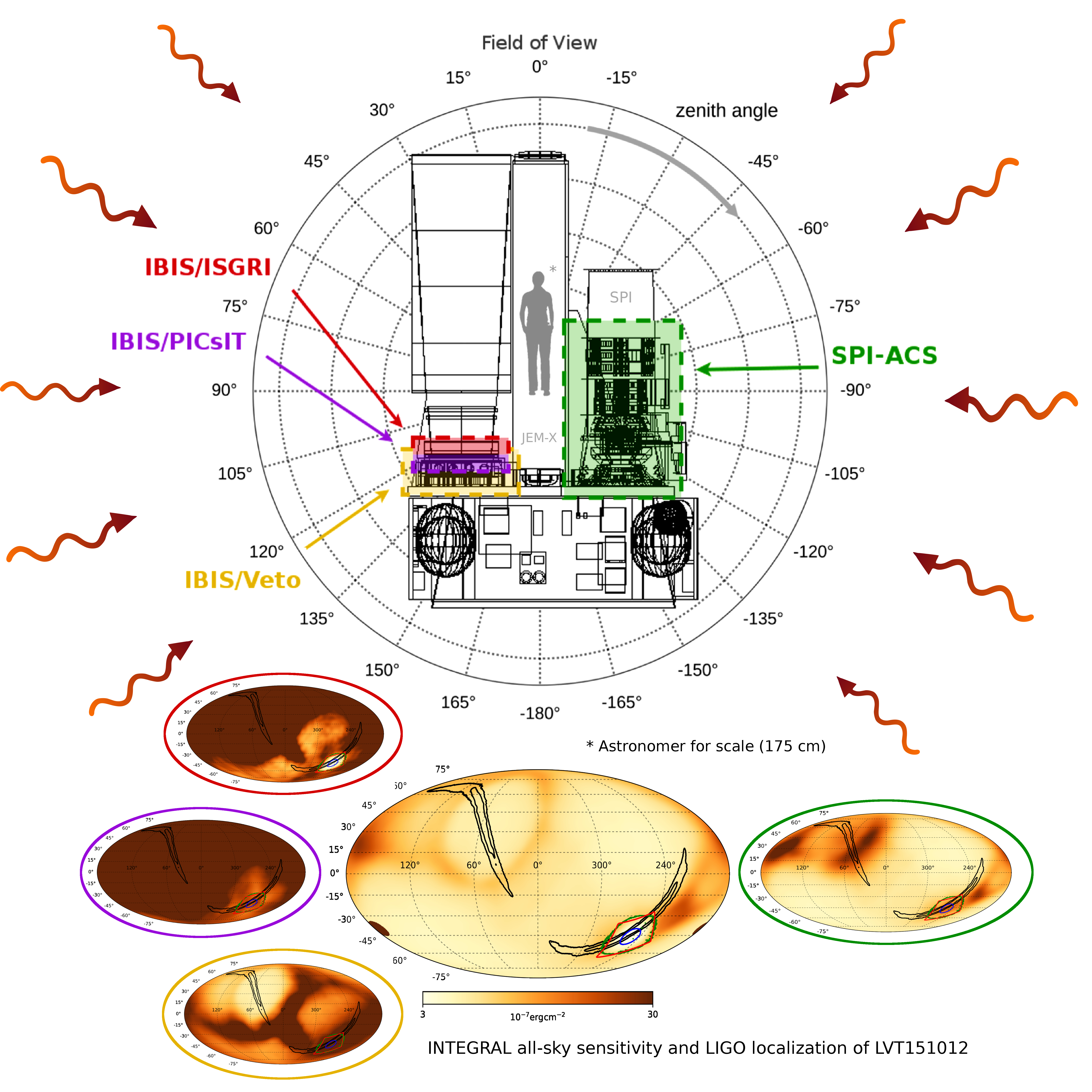}
	\caption{INTEGRAL's unique payload is comprised of a set of large and heavy detectors with sensitivity optimized in the field of view for SPI (grey), IBIS/ISGRI (red) and IBIS/PicSIT (violet), but extended marginally to the whole sky. Particle and radiation shields of the spectrometer (SPI-ACS in green) and of the imager (IBIS-Veto in yellow) are truly sensitive to the full sky  at a level
competitive with currently flying gamma-ray burst monitors. Coupled with an exceptionally stable background owing to the elongated orbit mostly far from the Earth's violent magnetosphere, this makes it an ideal instrument to search for electromagnetic counterparts to sources of various impulsive transients.
	The localization region of GW151012, shown in the bottom sky map, was very elongated and with a large fraction in the Field of View of the most sensitive INTEGRAL instruments at the time of occurrence of the LIGO/Virgo event. This permitted JEM-X, IBIS, and SPI measurements to be joined together, in order to derive a high sensitivity over more than 3 decades in photon energy: from 3 keV to 10 MeV.
		Observation of the complete, extended, localization region could only be achieved by combining the complementary contributions of both the INTEGRAL high-energy detectors and their active shields. This allowed the INTEGRAL team to derive the most stringent upper limit in a truly all-sky observation, constraining the ratio of energy released in gamma-rays to the gravitational wave energy to less than $4.4\times 10^{-5}$ (Adapted from \citealt{Savchenko2017}). }
\end{figure*}

Besides this routine search for GRBs, \intg can be exploited for targeted investigations of impulsive events to
find a time-coincident signal in gamma-rays. There are currently three types of events for which such a targeted search is performed:
gravitational waves (GW), high-energy neutrinos, and Fast Radio Bursts (FRBs). Target of Opportunity Observations (TOOs) can be performed to
follow-up a region of the sky in which an event is localized in a quest for a decaying gamma-ray source. The relatively large field of view of the imager permits regions of several hundred square degrees to be covered, reaching a depth of a few millicrab within one satellite revolution (lasting 2.7 days).

Gamma-ray bursts are routinely used for a triggered search of gravitational wave signals, so far unsuccessfully \citep{Abbott2019},
with the exception of the joint signal on GW180817/GRB170817A \citep[][and Sect.~\ref{Sec:GW170817}]{Abbott2017a}. Core-collapse supernovae are
also candidates for GW signals, but the GW detector sensitivity
would be currently enough only for a Galactic event  \citep{Abbott2016}. For instance, \intg was used to tentatively localize GRB051103 in M81 and the absence of a GW signal was exploited to suggest a giant magnetar flare as the
cause of this GRB \citep{Abadie2012}.

The LIGO and Virgo collaboration launched a call for partners to search for electromagnetic counterparts of gravitational wave events in early 2014 and the \intg team responded swiftly. A memorandum of understanding was signed to be informed in real time of alerts issued by the interferometers, with the constraint of keeping this information confidential. Any possible electromagnetic counterpart was to be shared with the LIGO-Virgo team and the other partners.
This agreement became effective on September 12, 2015, with the start of the first observing run (O1) of advanced LIGO \citep{Aasi2015}, which lasted until January 16, 2016, and for the second observing run (O2) from November 30, 2016 until August 25, 2017 \citep{Abbott2019a}.

Starting from the third run (O3) on April 1, 2019, GW alerts became public and more frequent, owing to the increased sensitivity \citep{Abbott2018}, requiring an expansion and reorganization
of the \href{https://www.astro.unige.ch/cdci/integral-multimessenger-collaboration}{INTEGRAL multi-messenger team}, besides an automation of analysis procedures. Members of the team have submitted proposals to reserve data rights and/or perform gamma-ray follow-up observations of GW event regions
in the \intg announcements of opportunity since cycle 13 in 2015.
Our team provided measurements of the gamma-ray flux for each event for which data are available (see Sect.~\ref{sec:other_events}). \intg's high duty cycle and all-sky detectors led to the seminal detection of a short GRB
due to the binary neutron star merger GW 170817 \citep[see Sect.~\ref{Sec:GW170817}][]{Savchenko2017b}.

Alerts from neutrino events are also public for IceCUBE\footnote{See the \href{https://icecube.wisc.edu/science/data/realtime_alerts}{Icecube alet system}.} \citep{Aartsen2017}, while an agreement has been put in place with the ANTARES experiment \citep{2011NIMPA.656...11A} to react to possible events and share results. Dedicated \intg time for follow-ups is reserved through accepted proposals.
\intg took part to the multi-wavelength campaign on the blazar TXS 0506+056 (3FGL J0509.4+0541) positionally consistent with the neutrino IC170922A
\citep{IceCubeCollaboration2018} (see Sect. \ref{Sec:PhysNu}).

Fast radio bursts (FRB) are currently one of the most mysterious phenomena in astronomy. They are sources emitting short ($\sim1-10$ ms) pulses of radio emission with peak fluxes of $\sim0.1-100$ Jy at 1.4 GHz, and  dispersion measures (DM) in excess of the Milky Way values along their lines of sight.
Together with their association with galaxies at cosmological distances, this points towards an extragalactic origin  (see \citealt{Cordes2019,Petroff2019} for reviews), but their origin remains elusive. Although connecting to multi-wavelength astrophysics, rather than multi-messenger, the search for hard-X-ray counterparts of these sources belongs
naturally to this review from the methodological point of view (see Sect.~\ref{Sec:PhysFRB}).

In the following, we review the \intg contribution in these fields, which gave a renewed science case for the mission in its late phase.

\section{Binary mergers and observable counterparts}
\label{Sec:PhysMergers}
Compact binary coalescences (CBCs) of black holes and neutron stars are among the loudest sources of gravitational waves in the current frequency window (from 15 Hz to a few kHz) of ground based interferometers, which are sensitive to the final part of their inspiral, merger and ringdown \citep{Abbott2019a}.

Three observing runs have been carried out completely by LIGO and Virgo: O1 from September 12, 2015 until January 19, 2016; O2, which started on November 30, 2016 and ended on August 25, 2017; and O3, which started on April, 1 2019 and ended on March, 27 2020 with a one-month commissioning break during October 2019. During the first two runs, the horizon for binary neutron star (BNS) mergers was limited to 100\,Mpc for LIGO Livingstone; 70--80\,Mpc for LIGO Hanford, and 30 Mpc for Virgo.
Thanks mainly to increased laser power, a squeezed vacuum source at the interferometer output,
and mitigating noise arising from scattered light,
the Livingston instrument began the O3 run with an average BNS range of 130 Mpc and
the Hanford instrument typically operated with an average range of 110 Mpc. Advanced Virgo reached
a BNS range of 50\,Mpc at the beginning of O3 \citep{Abbott2018}.


\subsection{Binary black hole mergers}
\label{Sec:PhysBBH}
Binary black-hole mergers involve only gravitational fields, as mass is all contained within the event horizon and cannot transmit any information outside. The merging event emits in the form of gravitational waves which can release a few solar masses of equivalent energy \citep[$3.0\pm0.5\,M_\odot c^2$ in GW 150914][]{Abbott2016}. The early report by the Fermi-GBM team of a gamma-ray signal possibly associated with GW150914 \citep{Connaughton2016,Connaughton2018}
triggered physicists to explore peculiar scenarios in which gamma-rays can be produced. These
include those BH-BH systems merging in very dense environments \citep[as for example in an AGN disk][]{Bartos2016}, or with dormant accretion disks \citep{Perna2016}, or even residing within an exploding star \citep{Loeb2016}. Since then, these scenarios have been challenged by the non-detection of any gamma-ray signal associated with BH-BH mergers. An extensive study with Fermi-GBM of the first LIGO Virgo catalog is reported by \citet{Burns2019}, this includes sub-significant triggers in LIGO-Virgo.
INTEGRAL observations were available for 20 out of the 25 events detected in O1 and O2 \citep{Abbott2019a}.
This is consistent with the INTEGRAL duty cycle of
about 85\%. In particular, observations are available for 7 out of 11 of
the high-confidence gravitational wave events and 13 out of 14
of the marginal ones.
For each of the observed events, INTEGRAL was sensitive to the entire
LIGO/Virgo localization region. A preliminary search reported in  \citet{2018GCN.23517....1S} did not reveal
any new significant impulsive gamma-ray counterparts, setting typical
upper limits on the 1-s peak flux ranging from $10^{-7}$ to $10^{-6}\mathrm{erg\,cm^{-2}\,s^{-1}}$
in the 75--2000\,keV energy range.
An investigation of the LIGO-Virgo O3 events reported in circulars
is detailed in Table~\ref{tab:gcns} and gives upper limits of similar values.
It is worth noting that an electromagnetic counterpart to a BBH merger with asymmetric masses
was reported as an optical luminosity variation in the accretion disk of an active galactic nucleus
50 days after the gravitational wave event S190521g by the Zwicky Transient Factory \citep{Graham2020}.
This is argued to be due to a Bondi accretion tail caused by the
kick velocity induced by the merger.

\subsection{Binary neutron star mergers}
\label{Sec:PhysBNS}

Tight binaries of neutron stars (NS) undergoing an in-spiral process and final merger, have long been of great interest in astrophysics as they represent a unique laboratory to investigate several long-standing questions. For a long time, such systems have been predicted to be the progenitors of short gamma-ray bursts \citep[sGRB,][]{nakar07} and the most promising sites for the production of heavy elements through the rapid neutron-capture process \citep[r-process,][]{freiburghaus99}. Moreover, they can be a useful tool to derive cosmological parameters, to investigate fundamental physics and to constrain the NS equation of state \citep[][]{Abbott2017e, Abbott2017, Abbott2017a,bauswein17,coughlin18}.

Binary neutron-star mergers
are the most promising sources from ground-based gravitational wave (GW) detectors to be detected in
the electromagnatic (EM) domain \citep[][]{nissanke13}.
A bright flare of gamma-rays was predicted to be followed not only by the
typical afterglow, but also by ultraviolet, optical, and infrared radiation coming from the reprocessing of nuclear decay products in the ejecta,
the so-called kilonova \citep[][]{kasen13,mezger14}.
These predictions were mostly confirmed in the case of the event observed on 17 August 2017, that we describe in detail in Sect~\ref{Sec:GW170817}.

If the product of the merger were a stable NS, several models would predict bright isotropic X-ray emission at different times.
Potentially powerful, nearly-isotropic, emission is expected if a NS-NS merger produces a long-lived millisecond magnetar.
In this case, X-ray to optical transients can be powered by the magnetar spin-down emission reprocessed
by the baryon-polluted environment surrounding the merger site (mostly due to isotropic matter ejection in the early
post-merger phase), with time scales of minutes to days and luminosity in the range $10^{43}-10^{48}\,\mathrm{erg\,s^{-1}}$
\citep[e.g.][]{Yu2013,Metzger2014,Siegel2016,Siegel2016a}.
However, in the most optimistic models, these transients can be detectable from minutes to hours after the event.
According
to alternative models, X-ray emission may also be generated via direct dissipation of magnetar winds \citep[e.g.][]{Zhang2013,Rezzolla2015}.
Furthermore, the high pressure of the magnetar wind can in some cases accelerate the
expansion of previously ejected matter into the interstellar medium up to relativistic velocities, causing a forward shock,
which in turn produces synchrotron radiation in the X-ray band\citep[with a high beaming factor of $\sim$0.8; see, e.g.][]{Gao2013}.

\citet{Sun2017} developed a detailed model for the X-ray post-merger emission from BNS mergers in the case where a long
lived NS is created. Of course, the outcome of a BNS merger depends on the Equation of State. Although
in most scenarios, NS with masses larger than 2.5M$_\odot$ will not survive as a stable body, numerical simulations have shown that a newly born millisecond magnetar may be created \citep[e.g.][]{Giacomazzo2013a,Gao2016}.
This magnetar scenario is supported by the existence of X-ray plateaus in the afterglows of some short GRBs.

Merger ejecta can represent $10^{-1}-10^{-3}\,M_\odot$ \citep[e.g.][]{Rezzolla2010,Rosswog2013}, and these ejecta cover
a significant part of the solid angle. In this case, the X-rays produced by internal dissipation within the
magnetar wind cannot escape freely, and this is called the ``trapped zone''. However they can heat and accelerate the
ejecta and eventually escape when the ejecta become optically thin at later times. \intg has the potential
to detect such radiation in rapid follow-up observations of these events, as discussed in Sect.\ref{Sec:follow-up}.

\section{High-energy neutrinos}
\label{Sec:PhysNu}
Produced in inelastic photo-hadronic (p$\gamma$) or hadronuclear (pp) processes, high-energy neutrinos in the
TeV--PeV range are the smoking-gun of hadronic interactions and cosmic-ray acceleration. Contrary to charged cosmic
rays, they are not deflected by magnetic fields and they do not suffer from absorption by pair production as do
high-energy photons.

A diffuse flux of high-energy neutrinos was discovered from 2013 by the IceCube experiment at the level of
$E^2_\nu F_\nu(E_\nu) \sim 10^{-8}~\textrm{GeV cm}^{-2}~\textrm{s}^{-1}~\textrm{sr}^{-1}$ per neutrino flavor
in the energy range between a few tens of TeV and a few PeV (see e.g. \citealt{IceCube2013}) but the sources
are still unknown. While the angular distribution of the astrophysical events is compatible with an isotropic
distribution, which favors an extragalactic origin, a sub-dominant contribution from Galactic sources is not
excluded. Multi-messenger astronomy, making use of neutrino, electromagnetic and/or gravitational wave signals
provides an increased discovery potential and good background reduction by looking for coincident detections
both in space and time. The good pointing accuracy of neutrino telescopes ($\lesssim$ 1 degree in the
muon-track channel) allows for fast electromagnetic follow-ups which are of primary importance to locate the
high-energy neutrino sources, in particular in the case of transient or variable ones.

The ANTARES \citep{2011NIMPA.656...11A} and IceCube \citep{2009NIMPA.601..294A} detectors are currently the
largest neutrino telescopes in operation respectively in the Northern and Southern hemispheres. By constantly
monitoring at least one complete hemisphere of the sky, they allow for complementary coverage with an almost
100\%-duty cycle, and thus are well designed to detect transient neutrino sources. Both telescopes operate
extensive programs of nearly real-time multi-wavelength (from radio to gamma-rays) follow-up
\citep{Ageron2012, Aartsen2017} as soon as a high-signal neutrino event is detected.

While such programs have not yet provided significant evidence for cosmic sources associated with HE
neutrinos, a few possible associations with active galactic nuclei (AGN) have already been claimed
\citep{2016NatPh..12..807K, 2017ApJ...843..109G,    2017ApJ...846..121L}. In particular, a compelling
case occurred in September 2017 when the LAT instrument on board Fermi and the MAGIC Cherenkov telescopes
observed enhanced gamma-ray emission from the BL Lac TXS 0506+056 (3FGL J0509.4+0541) positionally
consistent with the neutrino IC170922A \citep{IceCubeCollaboration2018}. The significance of this coincidence
between the blazar flare and the neutrino was evaluated to 3$\sigma$. Following this discovery, an analysis of
the archival neutrino data was performed by the IceCube collaboration which found a significant excess of neutrino emission
during 2014/2015 at the 3.5$\sigma$ level \citep{IceCubeCollaboration2018}. While intriguing, the
detection of a single neutrino does not allow unambiguous confirmation of the link between high-energy neutrinos
and blazars and more correlations will be needed to further assess the emission and particle acceleration
mechanisms. Likewise, previous cross-correlation studies using IceCube data showed that the population of
blazars observed by Fermi-LAT can only explain less than 20\% of the diffuse flux of astrophysical neutrinos
\citep{2017ApJ...835...45A}. While gamma-ray bursts (GRBs) are also disfavoured as a main contributor to the
diffuse flux, the question of the origin of high-energy neutrinos is a burning issue. Next-generation neutrino
telescopes such as KM3NeT (currently under deployment in the Mediterranean), IceCube-Gen2, foreseen by the end
of the decade at the South Pole and an upgrade of the Baikal neutrino telescope being built in Russia, have a
bright future ahead of them.

Regardless of the nature of the electromagnetic counterpart, multi-wavelength data are crucial to firmly identify the sources of high-energy neutrinos. In this context, INTEGRAL systematically follows ANTARES alerts, under an MoU agreement signed with the ANTARES collaboration and the IceCube triggers sent publicly through GCN notices \citep{Aartsen2017}. The INTEGRAL circulars are reported in Table~\ref{tab:gcns}.


\section{Fast radio bursts}
\label{Sec:PhysFRB}

Fast Radio Bursts (FRBs) consist of a single broadband pulse, with a duration of a few milliseconds and a flux density ranging typically from 0.1 to 30 Jy, but with the highest flux recorded so far of 120 Jy \citep{Ravi2016}.
They were discovered in wide-field pulsar surveys with the 1.4 GHz receiver at the Parkes radio telescope
\citep{Lorimer2007,Keane2012,Thornton2013}. The vast majority of FRBs was initially found by Parkes and ASKAP, but also Arecibo, Green Bank Telescope, and Molonglo have detected a bunch of them \citep{Spitler2016,Masui2015,Caleb2017}. At the time of writing, the CHIME experiment gives a major contribution \citep{CHIME/FRBCollaboration2019} with an \href{https://www.chime-frb.ca/repeaters}{online catalog of repeating FRB}. An earlier online FRB catalog \href{http://frbcat.org/}{frbcat.org/} collected events at least until late 2019 \citep{Petroff2016}.

FRBs have signals which are dispersed as in pulsars, but with much higher Dispersion Measures (DM, measuring the line-of-sight column density of free electrons).
Both a Galactic and an extragalactic origin have been initially proposed.
However, the large DM and the fact that they are preferentially found far from the Galactic plane
have favored an extragalactic origin.
Finally, their extragalactic nature has been firmly demonstrated by the association of a few FRBs with host galaxies
in the redshift range 0.0337--0.4755  \citep{Marcote2020,Prochaska2019}, and  possibly up to  z=0.66 \citep{Ravi2019}.
Two well characterized repeating FRBs (FRB121102, \citealt{Spitler2016,Scholz2016} and FRB 180814, \citealt{CHIME/FRBCollaboration2019})
made clear that at least some of them originate from non-destructive events, then eight more objects were reported by \citet{CHIME/FRBCollaboration2019} substantiating an
early hypothesis that there are two different populations of FRB, repeating and non-repeating \citep{Caleb2018}.
For FRB121102, repetitions allowed an association of the burst with a low mass galaxy at z = 0.19 \citep{Chatterjee2017,Marcote2017,Tendulkar2017}.
No X-ray and gamma-ray detection was found for this FRB \citep{Scholz2017}.
In some of the repeated bursts, sub-pulse frequency structure, drifting and spectral variation were reminiscent of that seen in FRB 121102,
suggesting similar emission mechanisms or propagation effects.
The increasing observational effort brought the identification of FRB180916.J0158+65 (in short FRB180916)
to be a repeating source with a characteristic periodicity of about 16 days \citep{CHIME/FRBCollaboration2020}, located in a star-forming region \citep{Marcote2020}.

The progenitors and the emission mechanisms responsible for the production of FRB are still unknown.
Brightness temperatures in FRBs are well in excess of thermal emission, requiring a
coherent emission process. For the case of repeating, extragalactic FRBs, giant pulses
from pulsars have been proposed \citep{Pen2015,Cordes2016}
as well as giant flares from magnetars \citep{Popov2013}. Explanations for extragalactic
sources of non-repeating FRB  (in the hypothesis that they are genuinely
one shot events and represent a different subclass of FRBs with respect to FRB121102) include evaporating primordial black holes \citep{Rees1977},
merging binary white dwarf systems \citep{Kashiyama2013}, merging neutron stars and white dwarfs \citep{Zhong2020},
merging neutron stars \citep{Hansen2001}, collapsing supramassive neutron stars \citep{Falcke2014}, and
superconducting cosmic strings \citep{Cai2012}.
\citet{Totani2013} has predicted that most short GRBs must be associated with FRBs in the NS-NS merging scenario and that a
longer time delay at lower frequencies may allow FRBs to be detected by follow-up searches after short GRBs.
For the interested reader, \citet{Platts2019} has compiled an extensive list of FRB models, which witnesses an extremely intense debate.
The localization of FRBs and characterization of their multi-wavelength counterparts
is essential to discriminate between the proposed models.
FRBs will also provide a new and powerful tool to probe the ``missing baryons'' component in the Universe.

The detection of hard X-ray flares and/or afterglows associated with an FRB would provide a
unique opportunity to study their progenitors and the associated radiative mechanisms.
INTEGRAL has already observed several among the phenomena proposed as arising from possible progenitors of FRBs.
For instance, giant magnetic flares have been observed \citep{Mereghetti2005,Mereghetti2009,Savchenko2010}.
When the satellite is promptly pointed to the region of interest, the principal instruments of INTEGRAL
have also detected faint long-lasting emission of GRBs  \citep{Martin-Carrillo2014}.
In principle, follow-up observations of the FRB error regions with INTEGRAL
could provide unique imaging capabilities with high sensitivity to lines (SPI) and accurate localization (IBIS).

\section{Methods for serendipitous search of impulsive events with \intg}
\label{Sec:INST}
As extensively described elsewhere in this series of reviews, \intg is equipped with
the `Spectrometer on INTEGRAL' \citep[SPI][]{Vedrenne2003,Roques2003},  the `Imager on Board the INTEGRAL Satellite' \citep[IBIS][]{Ubertini2003},
the `Joint European X-ray Monitors' \citep[JEM-X 1 and 2]{Lund2003}, and the
`Optical Monitoring Camera' \citep[OMC][]{Mas-Hesse2003}.
Here, we describe only the aspects of instruments and tools relevant for the detection and follow-up
of serendipitous events together with
the methods to efficiently combine the signals from different instruments.

\subsection{SPI}
The \intg SPI detector plane, made of 19 crystals of High Purity Germanium (GeD), has been designed to detect photons between 20 keV and 8 MeV and measure their energy with a precision ranging from $\sim$  2 to 5 keV over the whole energy domain.
Whether it is a prompt emission detection occurring by chance within the FoV or during a follow-up strategy, SPI gives access to fine spectroscopic information of the high energy properties for any potential GW counterpart. In particular, it will be able to measure or put upper limits for narrow emission lines, related to different physical mechanisms.
For instance, r-process elements are expected to be released during BNS mergers. The corresponding radioactive decays will produce  nuclear gamma-ray lines in the SPI energy domain. Similarly, pair production and annihilation physics  will be investigated in the 511 keV region. Typically, a 260 ks observation (2 INTEGRAL revolutions) provides a 3 $\sigma$ upper limit of $2.8 \times 10^{-4}$ ph cm$^{-2}$ s$^{-1}$ for a narrow annihilation line, between 505 and 515 keV.

\subsection{SPI-ACS}
The SPI is surrounded by an active
anti-coincidence shield \citep[SPI-ACS,][]{vonKienlin2003},
consisting of 91 (89 currently functional) BGO (Bismuth Germanate, Bi4Ge3O12) scintillator crystals.
The SPI-ACS is endowed with a large effective area
(up to $\sim$1\,m$^2$) $\gamma$-ray detector with a quasi-omnidirectional field
of view.
The ACS data are downlinked as event rates integrated over all the scintillator crystals with a time resolution of 50\,ms.
The typical number
of counts per 50 ms time ranges from about 3000 to 6000 (or
even more during high Solar activity).
A crucial property of the SPI-ACS data is that, contrary to
many other existing GRB detectors, the readout does not rely on any
trigger, so that a complete history of the detector count rate over
the mission lifetime is recorded. This opens the possibility of an offline search of GRBs or targeted searches.
The design of the ACS readout is such
that it provides almost no sensitivity to the direction of detected signals.
The SPI-ACS effective area and its
dependency on the direction and the energy is somewhat uncertain and
it can be investigated through detailed simulations of the photon propagation in the detector, as was done, for example by
\citet{Mereghetti2009}. However, this requires a mass model of
the entire INTEGRAL satellite. An alternative method involves making use of the events detected simultaneously by SPI-ACS
and other detectors. This approach was exploited by \citet{Vigano2009}.
A further development that combines both approaches has been pursued by
\citet{Savchenko2012,Savchenko2017}.

The SPI-ACS light curves are affected by the presence of short spikes ($\sim$50--150\,ms) that were
identified early on as cosmic-ray interactions by \citet{Rau2005} and confirmed as such by
\citet{Savchenko2012}.
In this work, it is also shown that the
decay of cosmic-ray induced radioactivity in BGO crystals, produces an ``afterglow''
discussed by \citet{Minaev2010} and mis-interpreted as a sign of a GRB nature.
The detailed knowledge of properties of the spikes
allowed \citet{Savchenko2012} to fully characterize the spikes and to separate them from the
real GRBs using a dedicated test statistic that is implemented in the targeted search of multi-messenger counterparts
as well as in post-processing of the events identified online by the IBAS system and
published online.

\subsection{IBIS/ISGRI}
ISGRI is the upper detector plane of IBIS \citep{Lebrun2003} and,
despite being optimized for imaging, it has some sensitivity also out of the field of view.
Indeed, the coded mask through which ISGRI usually observes the
high-energy sky cannot be fully exploited when searching for
impulsive events within the fully coded FoV when their location is not known.
While the sensitivity for a source in a fixed
location can be improved by using the coded mask pattern to
reject about 50\% of the background, this advantage is lost in
a search for a new source, when there are additional trial factors.
The conditions are different in the partially coded FoV, as a
progressively smaller fraction of the detector is exposed through
the coded mask holes and searches for short transients could be
optimized by considering a smaller portion of the detector relevant for specific directions.
This reduces the background, which
is proportional to the total effective area used for the search.
However, the instrument sensitivity is also reduced by exploring
lower effective areas, rapidly approaching that of the SPI-ACS
(see Figs.~\ref{fig:combined}).

We generally prefer to rely
on the light curves built from the entire detector to search for
impulsive events in the ISGRI data.
As the IBIS collimator tube becomes increasingly transparent at energies above $\sim$200\,keV, photons from directions that
are up to 80 deg off-axis with respect to the satellite pointing
can reach the ISGRI detectors, allowing this instrument to detect events occurring outside its FoV. Even soft events, with the
bulk of photons released below $\sim$200 keV, can be detected in ISGRI despite the absorption by the IBIS shield and other satellite
structures. For the very same reason, photons from soft events
produce a highly contrasted pattern on the ISGRI detector plane
that can be used to roughly constrain the source location (as is
done with a higher precision when the event is recorded through
the coded mask within the instrument FoV).

ISGRI is particularly well suited to searching for long transients i.e., those associated with GRB afterglows, because the
coded mask imaging allows us to better characterize the instrument background, as well as accurately subtract the contribution
of persistent sources from the data to probe the presence of
faint transients.

\subsection{IBIS/PICsIT}
IBIS/PICsIT \citep{Labanti2003} is the bottom detector layer of the IBIS telescope, located 90 mm below ISGRI.  It is composed of 4096 30 mm-thick CsI pixels (\( 8.2 \textrm{ mm} \times 8.2 \textrm{ mm} \)), featuring a total collecting area of about 2800 cm\(^2\) and is sensitive to photons between 175 keV and 10 MeV. In this energy range, the IBIS collimator tube is largely transparent and thus PICsIT can observe sources for all directions that are not occulted by the SPI instrument.  Some bright GRBs have been detected at angles of \( \sim 180^{\circ} \) from the satellite pointing direction.  The effective area of the instrument slowly  decreases as a function of off-axis angle, mainly due to the effect of the PICsIT planar geometry combined with the change in opacity of the shielding and ISGRI detector plane.

The instrument coded mask opacity to hard X-ray photons is larger than that of the passive shield thus PICsIT in principle collects more signal from sources outside the FoV than those closer to the satellite pointing direction (in sharp contrast with ISGRI).  This leads to an increased sensitivity for isolated, bright, impulsive events (i.e. GRBs). In these cases, the long-term background variability can often be neglected and its average level can be well-constrained before and after the event without relying on the coded mask.

To evaluate the response of PICsIT to high-energy bursts from any sky direction it is important to take into account the partial absorption of the corresponding radiation by the satellite structures.  We thus performed Monte Carlo simulations using the \intg mass model previously described by \citep{ferguson2003} and improving it through the inclusion of a more detailed IBIS mass model \citep{laurent2003}. We validated our approach by comparing the results for the detection of sources within the FoV with the predictions of the PICsIT responses based on the most recent instrument calibrations provided by the instrument team.

\subsection{IBIS-VETO}
The bottom and lateral sides of the IBIS detectors are surrounded by an active coincidence shield, the IBIS/Veto, which is made of 2-cm thick BGO crystals \citep{quadrini2003}.  The count rate of the IBIS/Veto is integrated continuously every 8 s and transmitted to ground.  This makes the subsystem an efficient detector of GRBs (and other gamma-ray transient phenomena) albeit with a reduced sensitivity for events shorter than the integration time.

We used Monte Carlo simulations exploiting the INTEGRAL mass model \citep{ferguson2003} to compute the IBIS/Veto response.  We checked our results by using observations of bright GRBs detected by \textit{Fermi}/GBM.  For a good match, we had to account for the low-energy threshold of the IBIS/Veto system for which we have a limited description.  The estimated discrepancy between the observed number of counts compared to those expected based on GBM results was found to be less than \( \sim 20 \)\%.

IBIS/Veto is a particularly useful instrument to study sources at off-axis angles larger than about \(120^{\circ} \) where the sensitivity of ISGRI, PICsIT, and the IBIS Compton mode are low.  At these angles, the coverage provided by SPI-ACS is also limited.  We also note that there is a relatively small fraction of the sky (about 15\%, depending on the source spectrum) for which the effective area of the IBIS/Veto is larger than the one of SPI-ACS.  For impulsive events longer than 8 s and near the opposite of the satellite pointing direction, the IBIS/Veto has a factor of 4 better sensitivity relative to the SPI-ACS due to its similar effective area (\( \sim 3000 \textrm{ cm}^2 \)), but lower background (by a factor of 2) and lower energy threshold.

The INTEGRAL/IBIS telescope is routinely used as a Compton Coded Mask telescope. True Compton scatterings are two events detected in the two IBIS independent detectors, ISGRI and PiCsIT. These IBIS/Compton data may be used to make Compton images out of the coded mask field of view. This is possible only at high energies, above 300 keV, when the IBIS shielding begins to become transparent. This Compton imaging process, which will be implemented in the INTEGRAL near real time transient follow-up system, will extend the IBIS FOV.

\subsection{JEM-X}
The Joint European Monitor for X-rays (JEM-X) instrument \citep{Lund2003} consists of a pair of coded-mask cameras providing a zero-response field of view of 13 degrees in diameter and an angular resolution of $3^\prime$.
With a position accuracy $\simeq 1^\prime$, JEM-X is especially useful to locate, and possibly identify, an X-ray counterpart if part of the localization region falls by chance in the field of view, or for pointed observations during follow-ups. Source detection can be achieved with a nominal continuum sensitivity of about $10^{-4}\,\mathrm{ph\,cm^{-2}\,s^{-1}\,keV^{-1}}$ from 3 to 20 keV for a 3$\sigma$ detection in $10^5$\,s, estimated at the beginning of the mission.

\subsection{OMC}
The Optical Monitoring Camera (OMC) was designed to observe the optical emission from the prime targets of the gamma-ray instruments on-board INTEGRAL \citep{mashesse2003}. It has a field of view of 5 by 5 degrees, but due to telemetry constraints only a set of preselected sources is transmitted to ground. This makes OMC unsuitable for a serendipitous search of impulsive events but it is a useful tool for pointed observations in follow-up campaigns, with a limiting V-Johnson magnitude in the range 16--17 depending on the sky direction. In case IBAS (see Sect.~\ref{Sec:IBAS}) localizes a GRB inside the OMC FoV, a telecommand is automatically sent to the satellite to set an appropriate CCD window in order to acquire OMC data at the GRB position. This happened just once on June 26, 2005, but the GRB happened so close to a bright star that the detector
was saturated.

\subsection{The IBAS software suite}
\label{Sec:IBAS}
The INTEGRAL Burst Alert System (IBAS) is the automatic
software devoted to the rapid detection and localization of GRBs
\citep{Mereghetti2003}. Contrary to many other $\gamma$-ray astronomy satellites, no on-board GRB triggering system is present
on INTEGRAL. Since the data are continuously transmitted
and reach the INTEGRAL Science Data Centre \citep[ISDC][]{Courvoisier2003} within a few seconds, the search for GRB is done at ISDC.
This has some advantages: besides the availability of larger computing power,
there is greater flexibility, with respect to systems
operating on board satellites, for software and hardware upgrades. To take full advantage of this flexibility, the IBAS software architecture
features different algorithms that are easily tunable using parameters.

IBAS localizations are based on two different programs using the data from
the IBIS lower energy detector ISGRI. The first program performs a simple monitoring
of the overall ISGRI counting rate. This is done by looking for
significant excesses with respect to a running average simultaneously on different time scales.
Excesses trigger an imaging analysis in which images are accumulated for different time intervals and
compared to the pre-burst reference in order to detect
the appearance of the GRB as a new source. This step is essential to eliminate many triggers due to instrumental effects and
background variations which do not produce a point source excess in the reconstructed sky images.
The second Detector Program is entirely based on image comparison. Images of the sky are continuously produced and compared with the previous ones to
search for new sources. This
one has the advantage of being less affected by variability of the
background or of other sources in the field of view.
Finally, a third kind of Detector Program is used to search
for GRBs detected by SPI-ACS.

Significant alerts are distributed using direct connections with partners who have subscribed
and transmitted by the GCN notice system\footnote{See \url{https://gcn.gsfc.nasa.gov/gcn/integral_grbs.html}}.
The potential of IBIS/ISGRI for the search of serendipitous events in the field of view is optimally exploited by IBAS with only a handful of GRBs not detected in the
online analysis, where they were identified as weak excesses \citep{Chelovekov2019}.

\subsection{An automatic system to react to transient events}
The results are the outcome of a well-defined process that assures a standard search is performed, as described below. In searching for impulsive events at the limit of instrumental sensitivity, it is essential that no ad hoc searches undermine the statistical robustness of the method. Otherwise, the estimate of association significance can be fooled and background events can appear as real. The human veto is introduced only to avoid obvious errors, due to unexpected technical issues, being propagated.
Given the crucial role of the pipeline, we describe it in some detail below.

As soon as a transient event, such as a neutrino or gravitational wave detection, is broadcast through a machine readable system, as a GCN notice, it is possible to automatically trigger and run a pipeline.
This INTEGRAL transient analysis pipeline features

\begin{itemize}
	\item  an initial assessment of the instrument status and the possibility to perform follow-up observations;
	\item a three part pipeline which starts with a realtime analysis of SPI-ACS data on a timescale of minutes, complements it with data  from other detectors on a time scale of one hour for a classification pipeline, and an extended untargeted search in SPI-ACS;
	\item complementary analysis of IBIS-Veto and PICsIT  spectral timing data.
\end{itemize}
These steps will be described in detail in a forthcoming paper.

In order to express the
upper limits and measurements in physical units, it is necessary to
make assumptions about the source spectra.
For the results of all pipelines, we use two spectral shapes characteristic of GRBs. The ``short-hard''
spectrum is close to an average short GRB spectrum detected by
Fermi/GBM, and is described by so-called ``Compton'' model with
$\alpha$=0.5, E$_{peak}$=600~keV \citep{Gruber2014}, used in conjunction
with 1~s timescale. The ``long-soft'' spectrum is suitable for long
bursts (typically thought to be associated to collapsar events), is a
Band GRB model, with $\alpha$=--1, E$_{peak}$=300~keV, $\beta$=--2.5.

It is only using these spectral templates, that we can compare the observations in different INTEGRAL all-sky instruments with the
predictions of the relative response model. Only with such a combination, is it possible
to find the small real variations in the stable INTEGRAL background to constrain long-lasting emission of GRBs and also to derive a crude localization.
In Fig.~\ref{fig:combined}, we show the range of detection significance for short (1 s) and long (8 s) GRB-like events as colored shaded regions. Out of the field of view, SPI-ACS is
the most sensitive instrument, but in the rear direction of the satellite, for long events, it is overcome by the IBIS/Veto system\footnote{The 8-s integration time of the IBIS-VETO signal prevents good sensitivity for short events}.
The different sensitivity of the instruments allows also for a very coarse localization of events: for instance if an event is strong in
the IBIS-VETO, but weaker in SPI-ACS, it is likely to come from the rear of the field of view. Whereas an event which is seen in SPI-ACS but also in the bare rates of ISGRI or PICsIT and not in the field of view  is likely to come from around 45 degrees from the pointing direction.
This coarse localization can range from 1\% to 75\% of the full sky,
depending on the spacecraft orientation with respect to the signal and on its spectral characteristics.
They are not usually suitable for afterglow follow-up, but they can be
combined with other constraints (e.g. from LIGO-Virgo, or from other
instruments).

Even though no public products are automatically distributed from the full
pipeline, example results may be found in \citep{Siegert2018,Margutti2019},
while the realtime products described in (ii) and (iii) are widely used in all GCN reports (see Table~\ref{tab:gcns}).

\section{Follow up observations}
\label{Sec:follow-up}
INTEGRAL is ideally suited to search for serendipitous gamma-ray signals, however, once a gravitational wave trigger is received, the large field of view of the pointed instruments can
be exploited to search for signals from the object formed during the merger.

In some cases, bright X- and gamma-ray emission is expected from BNS mergers. A
detection of such a bright X-ray counterpart would definitely point towards the
presence of a stable neutron star as the end-product of the merger \citep[e,g,][]{Metzger2014,Margalit2017}.
This is currently the most promising way to determine
whether the BNS merger product is a black hole or a neutron star and \intg has the potential to do so with its ability to cover several
hundreds of square degrees in a single dithering pattern.
In addition, radioactive decay is expected to produce
characteristic gamma-rays, which will leak out, and provide the most direct diagnostic of the kilonova energy source;
INTEGRAL limits for nearby events will be important \citep[see][for realistic limits]{Savchenko2017b}.
Early versus later time observations can constrain the viewing
angle and the geometry of the system, as well as the strength of the magnetic field and the rotation period of the newborn
neutron star.
On the other hand, while models for the EM emission from BBH merger are less developed, INTEGRAL upper limits
can be very valuable to constrain current and future theoretical efforts.

\section{Some selected results}
\label{Sec:Results}

\subsection{GW150914}
\label{Sec:GW150914}
The first detection of a gravitational wave signal in 2015 from the merger of two black holes with masses of about 30 solar masses each \citep{Abbott2016c}
was followed up by the first massive campaign of electromagnetic follow-up \citep{Abbott2016a,Abbott2016b}. No counterpart
was detected, although a lively discussion was raised by the tentative association with an excess in the Fermi-GBM detector count rates \citep{Connaughton2016}.
The significance of the event, based on the occurrence of similar excesses in the time series, is about $10^{-4}$, therefore, the probability that
such an event happens by chance at 0.4\,s from the GW trigger is about $3\times10^{-3}$, which implies an association significance of 2.9$\sigma$ \citep{Connaughton2018}.
The astrophysical origin of such an excess has been debated both from the perspective of the Fermi-GBM data analysis \citep{Greiner2016} and owing to the
non-detection of any excess in the SPI-ACS light curve, as reported by \citet{Savchenko2016}, who set an upper limit on the 75--2000\,keV fluence of $2\times 10^{-8}$\,erg\,cm$^{-2}$, equivalent to
about $10^{-6}$ the gravitational wave energy release.
As SPI-ACS was sensitive to the entire LIGO localization region, this upper limit put severe constraints
on the allowed spectrum of the Fermi-GBM excess: for an initially reported best-fit FERMI-GBN cutoff-power-law spectrum,
SPI-ACS would have detected a highly significant signal from 5 to 15$\sigma$ \citep{Savchenko2016}; for a power law without break, the signal would have been
much larger. \citet{Connaughton2018} acknowledged that measuring the spectrum of such a weak excess with Fermi-GBM alone has a high level of uncertainty and that
a deeper knowledge of the instrument cross-calibration with SPI-ACS would be beneficial to determine the allowed corners of the parameter space for which
the two signals are compatible.
In our view, the fact that no other excess was observed in the relatively large sample of gravitational wave signals from binary black holes \citep{TheFermiGamma-rayBurstMonitorTeam2020}
reinforces the conjecture that this excess was not of astrophysical origin.
Moreover, the only clear signal in both SPI-ACS and Fermi-GBM was due to a binary neutron star merger, as described in Sect.~\ref{Sec:GW170817}.

\subsection{GW170817}
\label{Sec:GW170817}

\begin{figure}
    \centering
    \label{fig:gw170817}
    \includegraphics[width=\columnwidth]{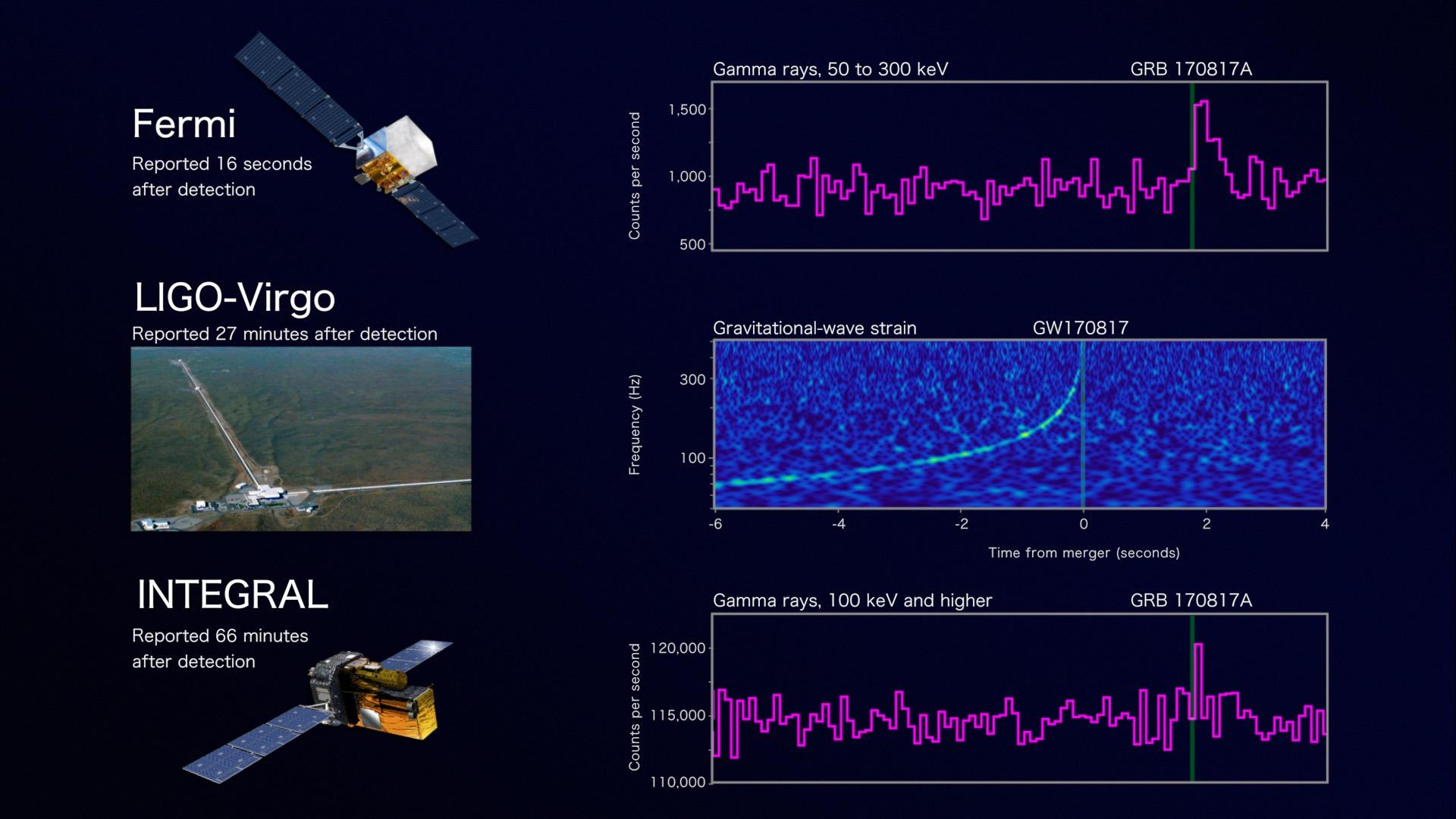}
    \caption{The joint, multi-messenger detection of GW170817 and GRB170817A. Top: The summed Fermi/GBM lightcurve for sodium iodide (NaI) detectors 1, 2, and 5 for GRB170817A between 50 and 300 keV, matching the 100 ms time bins of SPI-ACS data shown at the bottom. Middle: The time-frequency map of GW170817 was obtained by coherently combining LIGO-Hanford and LIGO-Livingston data. Bottom: The SPI-ACS lightcurve with the energy range starting approximately at 100 keV and with a high energy limit of least 80 MeV. All times here are referred to the GW170817 trigger time.}
\end{figure}

The loudest signal in gravitational waves so far was produced by the merging of two neutron stars in the galaxy NGC~4993 at 40~Mpc \citep{Abbott2017e}.
A gamma-ray burst was
autonomously detected by Fermi-GBM \citep{Goldstein2017} and independently reported by INTEGRAL SPI-ACS  with a fluence of \((1.4 \pm 0.4) \times 10^{-7} \textrm{ erg cm}^{-2}\)
in the \(75-2000\) keV energy range \citep{Savchenko2017b}. In Fig.~\ref{fig:gw170817}, we show the gravitational wave strain and the gamma-ray burst. The chance coincidence
of these events is $5\times 10^{-8}$, while
the time difference with the gravitational wave signal was $1.74\pm0.04$\,s \citep{Abbott2017a}. Such a short delay 1) constrains the difference
between the speed of light and gravity to
$$
-3 \times 10^{-15} < \Delta v / v < 7 \times 10^{-16}
$$
\noindent 2) places new bounds on the possible violation of Lorentz invariance
with a significant improvement on various parameters,
3) presents a new test of the equivalence principle constraining the Shapiro delay between gravitational and electromagnetic radiation.
This event marked the birth of the multi-messenger astronomy \citep[][]{Abbott2017} and enlightened the power of the multi-messenger approach.

Furthermore, GRB170817A is the closest gamma-ray burst detected so far
and is 100--1 million times weaker than any other events with a known distance.
Thus, it is consistent with being an off-axis GRB. The evolution of the gamma-ray emission towards a thermal
spectrum and the timescale of such an evolution are compatible with the scenario in which the GRB emission originated from the interaction of the jet with an
envelope of matter produced during the merger \citep{Goldstein2017}.
The absence of a bright hard X-ray counterpart \citep{Savchenko2017b} is compatible with the creation of a
black hole as a product of the merger.
Considering that the total mass of the GW170817 binary system is relatively large (2.74
M{$_\odot$} ), it is generally proposed that the merger of GW170817 would lead to a temporal hyper-massive neutron star (supported by
differential rotation) which survived 10--100 ms before collapsing into a BH or even directly a BH \citep{Margalit2017,Bauswein2017,Rezzolla2018,Metzger2018}.

After the serendipitous detection of the prompt GRB, INTEGRAL continued the planned observation for about 20 h.
Then, it was re-pointed, to perform a targeted TOO follow-up observation for several days.
This provided the most stringent upper limits on any electromagnetic signal in a very broad energy range, from 3 keV to 8 MeV. These data constrained the soft gamma-ray afterglow flux to \(< 7.1 \times 10^{-11} \textrm{ erg cm}^{-2} \textrm{ s}^{-1}\) in the range \(80-300\) keV.
In addition, it constrained the gamma-ray line emission intensity from radioactive decays, expected to be the principal source of the energy behind a kilonova event following a NS-NS coalescence. Finally, it provided a stringent upper limit on any delayed bursting activity, for example, from a newly formed magnetar. The gamma-ray prompt detection, and the subsequent continuous observations at all wavelengths, have provided important constraints on the high energy emission of the resulting kilonova and the nature of the post-inspiral object: NS, BH, or a new exotic object, still under debate.

An X-ray afterglow was detected only nine days after the merger, while the radio afterglow other seven days later \citep{Abbott2017}; this is the expected behavior of an ``orphan afterglow''.
In a standard GRB, the jetted emission is pointed towards us and the afterglow is seen immediately after the prompt emission; if there is misalignment,
the jet needs to open up before becoming visible.
The very low gamma-ray luminosity of GRB170817A implies that the jet was not seen on-axis or, in principle, that it was not even produced.
However, observations of the radio afterglow with the Very Long Baseline Inteferometer 160 days after the merger provided an estimation of the expansion speed \citep{Mooley2018,Mooley2018a,Mooley2018b,Ghirlanda2019}
and secured the existence of a standard jet as the origin also of the X-ray emission \citep{Troja2018,Margutti2018,D'Avanzo2018}. X-ray emission is still detected
almost three years after the event \citep{Troja2020}.
The off-axis angle inferred from gravitational waves is better constrained if the distance is fixed to the known distance of the host galaxy, by reducing parameter degeneracy and is
less than 28 degrees \citep{Abbott2017}.
Finally, the very low gamma-ray emission could also be produced by a structured cocoon around the jet, a phenomenon seen for the first time
in this object owing to its relative proximity \citep{Abbott2017a,Mooley2018}.

GW170817 was observed in gravitational waves by three detectors, this yielded a relatively small sky position area ($\sim$30
sq-deg), a good signal-to-noise and a precision of $\sim$25\% in the estimation of the distance of the source.
This allowed optical astronomers to restrict the pool of possible target galaxies for their search \citep{Abbott2017}.
Indeed, after 10 h from the GW signal detection, an optical counterpart was associated to GW170817 and
localized in the galaxy NGC 4993 at $\sim$40 Mpc by multiple and independent observers. The emission was
followed-up with multi-wavelengths observations (from UV to IR) for several weeks. The time-scale and the
color evolution of the light curve (from blue to red in few days) were compatible with a kilonova (KN)
scenario, in which the decay of heavy elements synthesized through r-process radioactively powered such
emission \citep[][]{kasen13,mezger14,pian17,smartt17}.
In particular, it has been explained as originating from two different ejecta components. A fast ($v \sim$0.3c)
dynamical ejecta, emitted from the polar regions is responsible of the early-time blue emission. This
component is characterised by a relatively high electron fraction, which is a signature of the occurrence of
weak interactions, triggered by the presence of a strong neutrino emitter, such as a hyper massive neutron
star \citep[HMNS][]{evans17}. Instead, the late-time red emission is dominated by lathanide-rich ejecta,
likely originating from an accretion disk wind in addition to an equatorial tidal ejecta \cite[][]{metzger17}.

\subsection{Other GW events}
\label{sec:other_events}
The response of all instruments can be combined to search for a signal and, in the majority of cases, obtain the most stringent upper limit on the electromagnetic signal in the INTEGRAL band.
This was shown for the INTEGRAL observation of GW151012 \citep{Savchenko2017}, which was not announced as an online alert, but after several months in their catalog paper \citep{Abbott2019a} in which it interpreted as a merger of two black holes of $23^{+18}_{-6}\mathrm{M}_\odot$ and $13^{+4}_{-5}\,\mathrm{M}_\odot$ at a distance of $1000\pm500$\,Mpc. In that case, the large localization region intersected also the field of view of the \intg instruments, which could exploited in their full dynamical range of sensitivity to derive upper limits on emission from 3 keV to 2 MeV.

Having multiple facilities able to detect gamma-ray impulsive events at the same time, albeit with different sensitivities and sky covering fractions, proved to be a fundamental asset also in the case of GW170104 for which INTEGRAL provided a stringent upper limit on the whole localization region \citep{Savchenko2017a}. The upper limit was incompatible in most of the sky with a marginal detection by the mini-Calorimeter (MCAL) onboard the AstroRivelatore Gamma a Immagini Leggero (AGILE). In the only limited portion of the sky where the sensitivity of the INTEGRAL instruments was not optimal and the lowest-allowed fluence estimated by the AGILE team would still be compatible with the INTEGRAL results, simultaneous observations by Fermi/Gamma-ray Burst Monitor and AstroSAT excluded an astrophysical origin of the AGILE excess.

During the third observing run of LIGO and Virgo, the INTEGRAL multi-messenger team and other groups have constantly monitored the instruments to look for
serendipitous signals. We report in Table~\ref{tab:gcns} all our circulars on this topic. In this sample, there was just one other very probable binary
neutron star merger  (GW 190425; \citealt{Abbott2020}), which gave origin to some controversy over a possible marginal detection with INTEGRAL between
\citep{Pozanenko2020} and the multi-messenger team \citep{Savchenko2019}.
The former report a weak gamma-ray burst in SPI-ACS consisting of two pulses $\sim$0.5 and $\sim$5.9 s after the NS merger
with an \emph{a priori} significance of 3.5 and 4.4$\sigma$.
Analysis of the SPI-ACS count rate history recorded  for a total of $\sim$125\,ks of observations around the event
has shown that the rate of random occurrence of two close spikes with similar
characteristics is such that a similar event occurs by chance, on average, every $\sim$4.3 hours.
The latter state that for the excess at 6 s after the GW trigger, they estimate a possible 75--2000 keV
fluence range due to uncertainty of the location
from 2$\times10^{-10}$ and  2$\times10^{-9}$ erg\,cm$^{-2}$ (in addition to systematic
uncertainty of response of 20\% and statistical uncertainty of 30\%),
assuming the duration of 1s and a characteristic short GRB spectrum with
an exponentially cut off power law with $\alpha=-0.5$ and $E_p=600$\,keV.
They stress that the FAP of association of this excess is below 3 $\sigma$.
However, no other counterpart was found
at any wavelength, mainly due to the relatively large distance of this object (160$\pm$70\,Mpc), but also
from the possibility that at least one of the components was a black hole.

\subsection{IceCube-170922A}
\label{Sec:IceCube}

On September 22$^{\textrm{nd}}$, 2017, at 20:54:30.43 (UTC), the IceCube neutrino telescope detected a high-energy muon track event (IC170922A) induced by a neutrino with an energy of $\sim$290~TeV with a 90\% confidence level lower limit of 183~TeV \citep{IceCubeCollaboration2018}. An automated alert notified the community 43~seconds later, providing preliminary position and energy estimates. Subsequent offline analyses led to a best-fitting right ascension of 77.43$^{+0.95}_{-0.65}$ and declination of +5.72$^{+0.50}_{-0.30}$ (degrees, J2000 equinox, 90\% containment region). Soon after this release, the neutrino was reported to be spatially correlated with the gamma-ray blazar TXS~0506+056, whose flaring episode was observed by Fermi-LAT \citep{2009ApJ...697.1071A} and by the MAGIC Cherenkov telescopes \citep{2016APh....72...76A}, up to about 400 GeV, within the following days \citep{IceCubeCollaboration2018}. Based on this correlation, a strong multi-wavelength follow-up campaign covered the full electromagnetic spectrum and allowed for an analysis of the broadband spectral energy distribution (SED) of TXS~0506+056. Assuming a redshift $z\sim0.34$ \citep{2018ApJ...854L..32P}, it was shown that the electromagnetic radiation of the blazar can be well explained by leptonic processes, with a radiatively subdominant hadronic component compatible with the detection of IC170922A (see e.g. \citealt{2018ApJ...864...84K} ).

The INTEGRAL observatory took part of the electromagnetic follow-up of this source at energies above 20 keV. Combining data from SPI-ACS and the veto of the IBIS imager, an upper limit on the 8-second peak flux at any time within $\pm$30 minutes from the alert time was estimated at the level of $10^{-7}$ erg~cm$^{-2}$~s$^{-1}$. From September 30$^{\textrm{th}}$ to October 24$^{\textrm{th}}$, the location of TXS~0506+056 was serendipitously in the field of view of INTEGRAL resulting to an effective exposure of 32~ks. The blazar was not detected in the ISGRI data and thus an upper limit on the average flux of $7.1\times10^{-11}$~erg~cm$^{-2}$~s$^{-1}$ and $9.8\times10^{-11}$~erg~cm$^{-2}$~s$^{-1}$ respectively in the energy range 20 keV -- 80 keV and 80 keV -- 250 keV was set (3$\sigma$ confidence level).

Even though those limits did not constrain the SED of TXS~0506+056, INTEGRAL was the only instrument able to cover the high-energy sky above $\sim$80~keV up to
the MeV range.
The energy range from tens of keV to tens of MeV is particularly interesting to constrain hadronic processes on the SED of blazars since relativistic protons interacting with synchrotron photons will produce secondaries whose synchrotron emission leaves an imprint in the energy range 40 keV -- 40 MeV \citep{2015MNRAS.447...36P}. In scenarios where the hard gamma-ray emission of blazars is produced by photohadronic interactions, the features of this process, also known as the Bethe-Heitler pair production process, may be comparable to the hard gamma-ray flux produced by photo-pion processes and thus can be an efficient way of constraining hadronic acceleration in blazar relativistic outflows. In this context, INTEGRAL and next-generation hard X-ray/soft gamma-ray instruments can play a crucial role in confirming the association between high-energy neutrinos and blazars.

\subsection{FRB counterparts}
\label{Sec:FRB}

In March 2018, three new FRBs were detected by the Parkes telescope (FRB180301 see ATel \#11376, FRB180309 see ATeL \#11385 and FRB180311 see ATeL \#11396).
For each event, the rates of the INTEGRAL ``all-sky detectors'' were searched for any impulsive transients at the time of the FRB \citep[as in][]{Savchenko2017}.
For the given FRB source location, the best sensitivity was achieved with IBIS/ISGRI or SPI ACS, depending on the source spectrum.
INTEGRAL did not detect any significant counterpart for these bursts, but set 3$\sigma$ upper limits for the 75-2000 keV
fluences of 4.0, 5.7 and 2.6e-7 erg\,cm$^{-2}$ for FRB180301 (ATeL \#11386), FRB180309 (ATeL \#11387) and FRB180311 (ATeL \#11431), respectively.

The periodic nature of FRB 180916 allowed observers to carry on targeted multi-wavelength campaigns: in one
of them, INTEGRAL provided a 3$\sigma$ upper limit on a
75--2000 keV fluence of any burst shorter than 1\,s (50\,ms) of
$1.8 \times 10^{-7}$ ($4 \times 10^{-8}$)\,erg\,cm$^{-2}$
\citep{Pilia2020}.
Unfortunately,  all three radio bursts found in the lowest radio frequency of the Sardinia Radio Telescope ($\sim$350 MHz)
occurred slightly more than one hour before the start of the
INTEGRAL pointing observation. Only the soft X-ray upper limit by XMM-Newton were available in correspondence of the burst
and they correspond to a limit in the $0.3-10$\, keV burst luminosity of
$\sim10^{45}$\,erg\,s$^{-1}$.
Similar results with radio activity not associated with X-ray flares were reported by \citet{Scholz2020}.

\begin{figure}
    	\centering
       \label{fig:frb-lc}
    \includegraphics[width=\columnwidth]{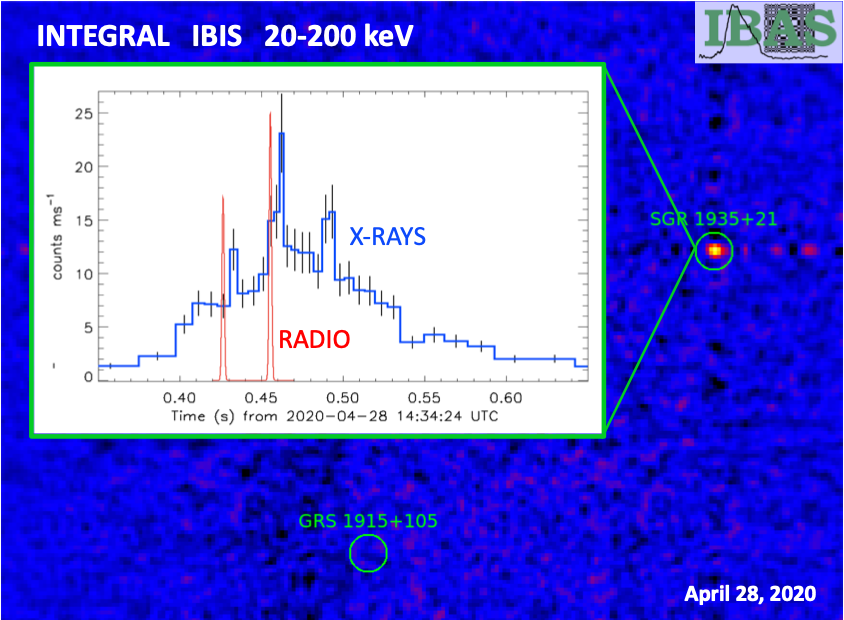}
    \caption{The localization of the source by the IBAS software and, in the inset, the
        background subtracted and deadtime corrected light curve derived from IBIS/ISGRI data in the 20--200 keV range.
        We used an adaptive binning to ensure at least 40 counts per time bin.   All the times are in the geocentric frame and referred to t$_o$=14:34:24 UTC of April 28, 2020. The red line (adapted from Fig.~1 of  \citealt{arxiv-CHIME}) marks the position of the radio pulses, represented with two Gaussian curves centered at 0.42648 s and 0.45545 s (adapted from \citealt{Mereghetti2020}).
    }
\end{figure}

A fundamental discovery has been made during an active period of the
galactic magnetar \mbox{SGR~1935$+$2154} in 2020 \citep{gcn27625,gcn27531}.
This culminated with the emission of  a ``burst forest'', i.e.  tens of bursts in  a short time interval on  April 27--28 \citep{atel13675,atel13678,gcn27659}.
INTEGRAL was observing the galactic black-hole binary GRS 1915+105, when IBAS detected two very intense bursts from the direction of the magnetar.
The brightest of the two was temporally coincident with  FRB 200428 discovered by the CHIME and STARE2  radio  telescopes  \citep{arxiv-CHIME,arxiv-STARE}. Its X-ray emission  was  detected also by instruments on the Insight-HXMT, Konus-WIND and AGILE satellites \citep{arxiv-HXMT,arxiv-konus,arxiv-AGILE}.
The background-subtracted and dead-time corrected light curve of the brightest burst, as measured with IBIS/ISGRI in the  20-200 keV energy range
is plotted in Fig.~\ref{fig:frb-lc}, together with the IBAS localization and the times of radio flares.
Spectral and temporal analysis revealed that this burst was not brighter than others, but it was
harder than the others detected by ISGRI and other satellites.
It has substructures superimposed to a general Gaussian-like profile. Two of these subpeaks
occurred just 6.5\,ms after the radio pulses and are separated by 30\,ms, exactly as the radio pulses.
The close time coincidence of the radio and X-ray emission could be due to common origin of both components in a relatively small region of the pulsar
magnetosphere (e.g., \citet{Lyutikov2002,Wadiasingh2019,Lyubarsky2020}). However, models involving emission at distances much larger than the light cylinder radius
($Pc/2\pi = 1.5\times10^{10}$ cm ) can produce (nearly) simultaneous pulses due to relativistic Doppler effects  \citep[e.g.,][]{Margalit2020a,Margalit2020}.
The observation of FRB-like radio emission and gamma-ray flares from a known galactic magnetar opens the possibility that a subset of
the currently known population of FRBs consists of galactic magnetars so far unidentified at other wavelengths,
while also providing strong support for a magnetar origin of extragalactic FRBs.

\subsection{Optical transient follow-up}
High-cadence optical surveys opened the possibility to follow-up, in X-rays, sources first detected
at longer wavelengths. New transients can
span large ranges in luminosity. They can be associated to H-stripped core collapse supernovae
or be extremes in supernova populations. There is a class of bright events outshining also supernovae
with blue colors the so called Fast Blue Optical Transients \citep[FBOT][]{Drout2014}.
Models that explain these phenomena span from the interaction of explosion shock waves with
circumstellar or atmospheric material to prolonged energy injection from
a central compact object.

AT 2018cow was discovered on 2018 June 16 by the ATLAS
survey, as a rapidly evolving transient located within a spiral arm
of the dwarf star-forming galaxy CGCG 137-068 at 60 Mpc.
\citet{Margutti2019} report on the first $\sim$100 days multi-wavelength follow-up
which uncovered, among others, a peculiar hard X-ray component
above $\sim$15\,keV, which outshone the underlying absorbed power-law
components, detected at soft X-rays 7.7 days after the discovery and rapidly
disappearing after another 3 days. The soft components continued to evolve
as $t^{-1}$ for more than a month.  This emission is larger than that seen in
supernovae and resembles GRBs in the local Universe. The
hard X-ray emission, whose cutoff at $\sim$50\,keV is uniquely constrained by INTEGRAL,
is interpreted as Compton reprocessing by a thick equatorial disk that
either shields the internal shocks or a central engine (like a magnetar).
At later stages, the rapid ejecta expansion causes clearance of scattering material and the disappearance
of the Compton hump.

This massive multi-wavelength follow-up campaign of an FBOT uncovered
a new class of astronomical transients powered by
a central engine and characterized by luminous
and long-lived radio plus X-ray emission.
The hard X-ray component played a central role in
the understanding of a Compton shield and motivates
continuation of this follow-up activity with INTEGRAL.

\section{Conclusions and future perspectives}

Nearly fifteen years after the \textit{INTEGRAL} scientific operation started, the space observatory entered a new exciting phase of its scientific life, playing a major role in the era of ``Multimessenger astrophysics''.
In fact, the highly eccentric orbit coupled with a set of complementary detector features, providing continuous coverage of the
whole sky, gives INTEGRAL unprecedented capability for the identification and study of the electromagnetic radiation associated
with multi-messenger signals.

INTEGRAL provided, in most of the cases, the best upper limits available to binary black hole mergers with a
ratio of emitted electromagnetic to gravitational energy $E_{\gamma}/E{GW} \sim < 10^{-7} - 10^{-5}$, clearly demonstrating
the absence of impulsive gamma-ray burst emission contemporaneous with GW.

The independent detection by \intg of the short gamma-ray burst GRB170817A 1.7s after the end of the GW signal,
has shown a completely different scenario in the case of NS-NS mergers,
proving its association with the binary neutron star merging event GW170817 detected by the LVC.
In fact, the association was immediately evident, due to the  time lag of the two signals and the positional coincidence,
with the overlapped error box, derived from gravitational waves, \intg/SPI-ACS and \textit{Fermi}/GBM,
reported almost in real time. The GW170817 detection from LVC and the corresponding detection of GRB170817A have been a fundamental step in multi-messenger astrophysics with a total combined statistical significance of 5.3 \( \sigma\) for the joint GW-GRB detection. It also firmly demonstrated the correlation between GW emission and the kilonova as a product of the NS-NS inspiral.

The 1.74s delay between the GW arrival time and the detection of gamma-rays, after a travel time of \( \sim 130\) million years also places strict limits between the speed of light and gravitational waves in the (not fully vacuum) universe with an unprecedented accuracy. Furthermore, the time delay of \((+1.74 \pm 0.05)\) s between GRB170817A and GW170817 implies new bounds on Lorentz Invariance Violation and revises the test of the equivalence principle by constraining the Shapiro delay between gravitational and electromagnetic radiation. Finally, we have used the time delay to constrain the size and bulk Lorentz factor of the region emitting gamma-rays \citep{Abbott2017a}.



The low luminosity and flux of GRB 170817A has suggested the possible existence of a population of short GRBs that are below instrument thresholds and are missed due to the lack of on-board trigger. Initial GBM results report detecting \( \sim 80\) SGRBs per year, compared to \( \sim 40\) triggered events per year. One of the ongoing and future activities will be to search for sub-threshold short GRBs in PICsIT and SPI-ACS data for untriggered events reported by GBM. This common search for past sub-threshold events can be used by the LIGO-Virgo collaboration to search for low-significance GW signals. During the next LIGO-Virgo observing runs, near real-time sub-threshold detections for the common search can be used to look for faint GW counterparts and lead to follow-up observations across the EM spectrum.
Finally, the production of quasi real-time PICsIT spectra, with a time resolution of 7.5 ms over 8 energy channels from 0.25 to 2.6 MeV is ongoing. The major difficulty is the production of a reliable deconvolution matrix taking into account the different azimuth and elevation angles of the detected burst. This will complete the existing real-time processing, already showing at the same time from each triggered or alerted GRB, the count rates measured by SPI-ACS, IBIS/PICSiT and IBIS/VETO.

\begin{figure}
	\centering
	\includegraphics[width=\columnwidth]{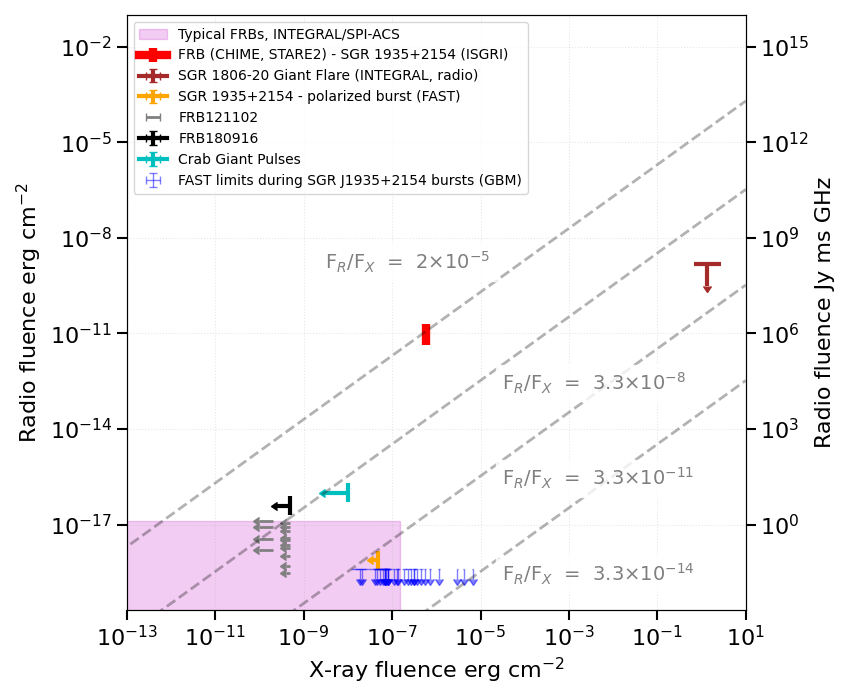}
	\caption{ \label{fig:FxFr} Radio versus  X-ray (starting from 0.5~keV)  fluences  for FRBs and magnetar bursts.
		The range of FRB fluences corresponds to a variety of detections reported in the past years \citep{Hurley2005,Tendulkar2016,Kozlova2016,Scholz2017,arxiv-FAST, arxiv-CHIME, arxiv-STARE,Scholz2020,Pilia2020,Karuppusamy2010}. The purple region indicates a robust upper limit on the hard X-ray fluence of FRBs as derived with a high-duty-cycle detector, such as the INTEGRAL SPI-ACS \citep[from][]{Mereghetti2020}.
	}
\end{figure}

\emph{Fast radio bursts} were thought to be linked to magnetar giant flares, but the association of a a Galactic FRB with
a gamma-ray flare of the magnetar SGR~1935$+$2154 constituted a game changer with a ratio of radio to X-ray luminosity of $2\times10^{-5}$.
Not only has this association confirmed such a long-sought association, but also
opened the interesting possibility that a fraction of FRBs could be of Galactic origin. On the other hand,
tight X-ray upper limits on periodic FRBs showed that often the X-ray luminosity must be at least three orders of magnitude less
than in the observed case. Radio upper limits on the Galactic magnetar flares are even tighter arriving to a range of radio to X-ray fluence values $\sim10^{-11}$ (see Fig.~\ref{fig:FxFr}). Despite this huge intrinsic variability, there is a huge discovery space open
to serendipitous discoveries or to targeted observations of repeating FRBs or magnetar flaring periods.
Indeed, not all FRB emit X-rays and not all magnetar flares emit radio waves, but we need to investigate the proposal by \citet{Mereghetti2020} that
spectral and timing characteristics of gamma-ray flares may be linked to radio emission.

The unique INTEGRAL performance discussed above are relevant also in the search for counterparts of astrophysical neutrinos, as demonstrated in several recent cases for which constraining upper limits were provided.

One lesson that can be learned from the INTEGRAL results described above, is that unanticipated uses of a payload can give important scientific contributions and exciting results. By definition, it is difficult to optimize the mission for an unforeseen science exploitation, but some general guidelines can be followed, as including the possibility of reconfiguration of the on-board software (with the associated problem of maintaining the required expertise for an extended time period). Also important are an accurate calibration of all the active elements (including unconventional directions and energies), as well as a complete characterization of both payload and spacecraft with an accurate mass model.

\section{Acknowledgements}
This work is based on observations with INTEGRAL, an ESA project with instruments and science data center funded by ESA member states (especially the PI countries: Denmark, France, Germany, Italy, Switzerland, Spain), and with the participation of Russia and the USA. The INTEGRAL SPI project has been completed under the responsibility and leadership of CNES. The SPI ACS detector system has been provided by MPE Garching/Germany. The SPI team is grateful to ASI, CEA, CNES, DLR, ESA, INTA, NASA, and OSTC for their support. The Italian INTEGRAL team acknowledges the support of ASI/INAF agreement No. 2013-025-R.1 and 2019-35-HH. R.D. and A.v.K. acknowledge the German INTEGRAL support through DLR Grant 50 OG 1101. A.D. acknowledges the support of the Spanish State Research Agency grants ESP2017--87676--C5--1--R (AEI/FEDER, UE) and MDM--2017--0737 (Unidad de Excelencia Mar{\'\i}a de Maeztu -- CAB). F.O. acknowledge the support of the H2020 European Hemera program, grant agreement No 730970. We acknowledge the continuous support by the INTEGRAL Users Group and the exceptionally efficient support by the teams at ESAC, ESOC, and ISDC for the scheduling and prompt elaboration of the targeted follow-up observations.




\bibliographystyle{elsarticle-harv}
\bibliography{savchenko,frb,gcnatel,gw,special}

\input{Table/gcntable.tex}

\end{document}

%% file: Table/gcntable.tex

{
	\small
	\onecolumn
	\appendix
	\section{The list of GCNs related to INTEGRAL results on multi-messenger astronomy}
	
	\renewcommand{\arraystretch}{1.5}
	\begin{longtable}{c|c|C{8.5cm}|C{2.2cm}}
		
	\caption{\label{tab:gcns}A collection of multi-messenger GCN
		circulars based on INTEGRAL data (the first column contains an hyper-link to the GCN). 
		The one flagged with $^\ddagger$ were published by teams other than
		the \href{https://www.astro.unige.ch/cdci/integral-multimessenger-collaboration}{INTEGRAL multi-messenger team}.
		The time of the trigger events or the circular submission time (flagged with $^\dagger$) is reported
		in the second column. 
		The $3\sigma$ 75--2000\,keV fluence upper limit refers to 
		impulsive events, not to follow-up observations. It is expressed in 
		erg\,cm$^{-2}$ and obtained using a combination of INTEGRAL detectors as described in \citet{Savchenko2017} within the
		50\% probability containment region of the source localization. The template model is a
		burst lasting less than 1 s with a characteristic short GRB spectrum
		(an exponentially cut off power law with $\alpha=-0.5$ and $E_p$=600 keV)
		occurring at any time in the interval within 300 s around the event time.}\\
	\hline
	\hline
	GCN  & Event time (UT) & GCN title &  3$\sigma$ 75--1000\,keV fluence u.l.\\
	\hline
	\endfirsthead

	GCN  & Event time & GCN title &  3$\sigma$ 75--1000\,keV fluence u.l.\\
	 \hline
	 \endhead

	\hline
	\multicolumn{4}{r}{\small\sl continued on next page} \\
	\hline 
	\endfoot

	\hline
	\endlastfoot

                            \href{https://gcn.gsfc.nasa.gov/gcn3/18354.gcn3}{GCN18354} 
                                	&
                				                   	2015-09-21 18:52:51$^\dagger$ & 
                                   LIGO/Virgo G184098: INTEGRAL search of temporally coincident prompt hard X-ray emission &

                                    $<$1.3$\times$10$^{-7}$ \\

                            \href{https://gcn.gsfc.nasa.gov/gcn3/18737.gcn3}{GCN18737} 
                                	&
                							2015-12-26 03:38:53 &
                                LIGO/Virgo G211117: INTEGRAL was inactive at the time of the event &

                                    -- \\

                            \href{https://gcn.gsfc.nasa.gov/gcn3/19789.gcn3}{GCN19789} 
                                	&
                				                   	2016-08-10 16:40:10$^\dagger$ & 
                                   INTEGRAL/SPI-ACS and IBIS/VETO search for prompt gamma-ray counterpart of IceCube-160806A &

                                    $<$2.2$\times$10$^{-7}$ \\

                            \href{https://gcn.gsfc.nasa.gov/gcn3/19828.gcn3}{GCN19828} 
                                	&
                							2016-08-14 21:45:54 &
                                INTEGRAL/SPI-ACS observation of IceCube HESE 128340 58537957 &

                                    $<$3.1$\times$10$^{-7}$ \\

                            \href{https://gcn.gsfc.nasa.gov/gcn3/20122.gcn3}{GCN20122} 
                                	&
                							2016-11-03 09:07:31 &
                                INTEGRAL SPI-ACS and IBIS/Veto observation of IceCube-161103 &

                                    $<$4.6$\times$10$^{-7}$ \\

                            \href{https://gcn.gsfc.nasa.gov/gcn3/20249.gcn3}{GCN20249} 
                                	&
                							2016-12-10 20:07:16 &
                                INTEGRAL SPI-ACS and IBIS/Veto observation of IceCube-161210 &

                                    $<$4.6$\times$10$^{-7}$ \\

                            \href{https://gcn.gsfc.nasa.gov/gcn3/20366.gcn3}{GCN20366} 
                                	&
                							2017-01-04 10:11:58 &
                                LIGO/Virgo G268556: INTEGRAL search of temporally coincident prompt hard X-ray emission &

                                    $<$1.6$\times$10$^{-7}$ \\

                            \href{https://gcn.gsfc.nasa.gov/gcn3/20496.gcn3}{GCN20496} 
                                	&
                							2017-01-20 12:30:59 &
                                LIGO/Virgo G270580: INTEGRAL search of temporally coincident prompt hard X-ray &

                                    $<$1.5$\times$10$^{-7}$ \\

                            \href{https://gcn.gsfc.nasa.gov/gcn3/20694.gcn3}{GCN20694} 
                                	&
                							2017-02-17 06:05:55 &
                                LIGO/Virgo G274296: INTEGRAL search for a prompt gamma-ray counterpart &

                                    $<$1.4$\times$10$^{-7}$ \\

                            \href{https://gcn.gsfc.nasa.gov/gcn3/20755.gcn3}{GCN20755} 
                                	&
                							2017-02-25 18:30:21 &
                                LIGO/Virgo G275404: INTEGRAL search for a prompt gamma-ray counterpart &

                                    $<$1.5$\times$10$^{-7}$ \\

                            \href{https://gcn.gsfc.nasa.gov/gcn3/20768.gcn3}{GCN20768} 
                                	&
                							2017-02-27 18:57:31 &
                                LIGO/Virgo G275697: INTEGRAL search for a prompt gamma-ray counterpart &

                                    $<$1.5$\times$10$^{-7}$ \\

                            \href{https://gcn.gsfc.nasa.gov/gcn3/20834.gcn3}{GCN20834} 
                                	&
                				                   	2017-03-07 22:38:53$^\dagger$ & 
                                   LIGO/Virgo G275404 and AGL J1914+1043: INTEGRAL follow-up &

                                    -- \\

                            \href{https://gcn.gsfc.nasa.gov/gcn3/20856.gcn3}{GCN20856} 
                                	&
                				                   	2017-03-12 21:21:29$^\dagger$ & 
                                   INTEGRAL observation of IceCube HESE 65274589 129281 &

                                    $<$4.6$\times$10$^{-7}$ \\

                            \href{https://gcn.gsfc.nasa.gov/gcn3/20867.gcn3}{GCN20867} 
                                	&
                							2017-03-13 22:40:09 &
                                LIGO/Virgo G277583: INTEGRAL was inactive at the time of the event &

                                    -- \\

                            \href{https://gcn.gsfc.nasa.gov/gcn3/20928.gcn3}{GCN20928} 
                                	&
                				                   	2017-03-21 23:03:41$^\dagger$ & 
                                   INTEGRAL SPI-ACS observation of AMON IceCube EHE 80305071 129307 &

                                    $<$1.9$\times$10$^{-7}$ \\

                            \href{https://gcn.gsfc.nasa.gov/gcn3/20933.gcn3}{GCN20933} 
                                	&
                							2017-03-13 22:40:09 &
                                LIGO/Virgo G277583: INTEGRAL search for an X-ray and gamma-ray counterpart &

                                    -- \\

                            \href{https://gcn.gsfc.nasa.gov/gcn3/20937.gcn3}{GCN20937} 
                                	&
                				                   	2017-03-25 18:40:15$^\dagger$ & 
                                   INTEGRAL pointed follow-up of IceCube-170321A &

                                    -- \\

                            \href{https://gcn.gsfc.nasa.gov/gcn3/21063.gcn3}{GCN21063} 
                                	&
                							2017-05-02 22:26:07 &
                                LIGO/Virgo G284239: INTEGRAL search for a prompt gamma-ray counterpart &

                                    $<$1.5$\times$10$^{-7}$ \\

                            \href{https://gcn.gsfc.nasa.gov/gcn3/21231.gcn3}{GCN21231} 
                                	&
                							2017-06-08 02:01:16 &
                                LIGO/Virgo G288732: INTEGRAL search for a prompt gamma-ray counterpart &

                                    $<$4.3$\times$10$^{-7}$ \\

                            \href{https://gcn.gsfc.nasa.gov/gcn3/21440.gcn3}{GCN21440} 
                                	&
                							2017-08-09 08:28:21 &
                                LIGO/Virgo G296853: no INTEGRAL data at the time of the event &

                                    -- \\

                            \href{https://gcn.gsfc.nasa.gov/gcn3/21476.gcn3}{GCN21476} 
                                 	$^\ddagger$ &
                				                   	2017-08-14 14:39:09$^\dagger$ & 
                                   LIGO/Virgo G297595: SPI-ACS/INTEGRAL data investigations &

                                    -- \\

                            \href{https://gcn.gsfc.nasa.gov/gcn3/21478.gcn3}{GCN21478} 
                                	&
                				                   	2017-08-14 16:13:19$^\dagger$ & 
                                   LIGO/Virgo G297595: INTEGRAL search for a prompt gamma-ray counterpart &

                                    $<$2.1$\times$10$^{-7}$ \\

                            \href{https://gcn.gsfc.nasa.gov/gcn3/21507.gcn3}{GCN21507} 
                                	&
                				                   	2017-08-17 13:57:47$^\dagger$ & 
                                   LIGO/Virgo G298048: INTEGRAL detection of a prompt gamma-ray counterpart &

                                    -- \\

                            \href{https://gcn.gsfc.nasa.gov/gcn3/21611.gcn3}{GCN21611} 
                                	&
                							2017-08-19 15:50:46 &
                                LIGO/Virgo G298389: INTEGRAL search for a prompt gamma-ray counterpart &

                                    $<$4$\times$10$^{-7}$ \\

                            \href{https://gcn.gsfc.nasa.gov/gcn3/21658.gcn3}{GCN21658} 
                                	&
                							2017-08-23 13:13:58 &
                                LIGO/Virgo G298936: INTEGRAL was inactive at the time of the event &

                                    -- \\

                            \href{https://gcn.gsfc.nasa.gov/gcn3/21672.gcn3}{GCN21672} 
                                	&
                				                   	2017-08-24 09:03:02$^\dagger$ & 
                                   LIGO/Virgo G298048: INTEGRAL pointed follow-up observations &

                                    -- \\

                            \href{https://gcn.gsfc.nasa.gov/gcn3/21695.gcn3}{GCN21695} 
                                	&
                				                   	2017-08-25 14:33:57$^\dagger$ & 
                                   LIGO/Virgo G297595: INTEGRAL pointed follow-up observations &

                                    -- \\

                            \href{https://gcn.gsfc.nasa.gov/gcn3/21699.gcn3}{GCN21699} 
                                	&
                				                   	2017-08-25 17:06:50$^\dagger$ & 
                                   LIGO/Virgo G299232: INTEGRAL search for a prompt gamma-ray counterpart &

                                    $<$3.6$\times$10$^{-7}$ \\

                            \href{https://gcn.gsfc.nasa.gov/gcn3/21917.gcn3}{GCN21917} 
                                	&
                				                   	2017-09-23 09:31:27$^\dagger$ & 
                                   INTEGRAL SPI-ACS observation of AMON IceCube-170922A &

                                    $<$2.8$\times$10$^{-7}$ \\

                            \href{https://gcn.gsfc.nasa.gov/gcn3/22018.gcn3}{GCN22018} 
                                	&
                							2017-10-15 01:34:30 &
                                INTEGRAL SPI-ACS observation of IceCube-171015A &

                                    $<$2.2$\times$10$^{-7}$ \\

                            \href{https://gcn.gsfc.nasa.gov/gcn3/22109.gcn3}{GCN22109} 
                                	&
                							2017-11-06 18:39:39 &
                                INTEGRAL observation of IceCube-171106A &

                                    $<$2.7$\times$10$^{-7}$ \\

                            \href{https://gcn.gsfc.nasa.gov/gcn3/22167.gcn3}{GCN22167} 
                                	&
                				                   	2017-11-24 21:08:15$^\dagger$ & 
                                   INTEGRAL pointed follow-up of IceCube-171106A &

                                    -- \\

                            \href{https://gcn.gsfc.nasa.gov/gcn3/23221.gcn3}{GCN23221} 
                                	&
                							2018-09-08 19:59:32 &
                                INTEGRAL observation of IceCube-180908A &

                                    $<$2.8$\times$10$^{-7}$ \\

                            \href{https://gcn.gsfc.nasa.gov/gcn3/23517.gcn3}{GCN23517} 
                                	&
                				                   	2018-12-07 11:09:45$^\dagger$ & 
                                   INTEGRAL observations of the events in the GWTC-1 catalog &

                                    -- \\

                            \href{https://gcn.gsfc.nasa.gov/gcn3/23689.gcn3}{GCN23689} 
                                	&
                							2019-01-04 08:34:38 &
                                INTEGRAL observation of IceCube-190104A &

                                    $<$2.1$\times$10$^{-7}$ \\

                            \href{https://gcn.gsfc.nasa.gov/gcn3/23807.gcn3}{GCN23807} 
                                	&
                				                   	2019-01-29 15:48:57$^\dagger$ & 
                                   INTEGRAL observation of IceCube-190124A &

                                    $<$5.1$\times$10$^{-7}$ \\

                            \href{https://gcn.gsfc.nasa.gov/gcn3/23927.gcn3}{GCN23927} 
                                	&
                							2019-02-21 08:25:40 &
                                INTEGRAL observation of IceCube-190221A &

                                    $<$5.2$\times$10$^{-7}$ \\

                            \href{https://gcn.gsfc.nasa.gov/gcn3/24066.gcn3}{GCN24066} 
                                	&
                				                   	2019-04-08 21:04:49$^\dagger$ & 
                                   INTEGRAL observation of S190408an &

                                    $<$2.5$\times$10$^{-7}$ \\

                            \href{https://gcn.gsfc.nasa.gov/gcn3/24101.gcn3}{GCN24101} 
                                	&
                							2019-04-12 05:30:44 &
                                INTEGRAL prompt observation of S190412m &

                                    $<$2.9$\times$10$^{-7}$ \\

                            \href{https://gcn.gsfc.nasa.gov/gcn3/24139.gcn3}{GCN24139} 
                                	&
                							2019-04-08 18:18:02 &
                                LIGO/Virgo S190408an: INTEGRAL follow-up observations &

                                    -- \\

                            \href{https://gcn.gsfc.nasa.gov/gcn3/24152.gcn3}{GCN24152} 
                                	&
                							2019-04-21 21:38:56 &
                                LIGO/Virgo S190421ar: INTEGRAL prompt observation &

                                    $<$1.9$\times$10$^{-7}$ \\

                            \href{https://gcn.gsfc.nasa.gov/gcn3/24169.gcn3}{GCN24169} 
                                	&
                							2019-04-25 08:18:05 &
                                LIGO/Virgo S190425z: INTEGRAL prompt observation &

                                    -- \\

                            \href{https://gcn.gsfc.nasa.gov/gcn3/24170.gcn3}{GCN24170} 
                                 	$^\ddagger$ &
                							2019-04-25 08:18:05 &
                                LIGO/Virgo S190425z: INTEGRAL SPI-ACS prompt observation &

                                    -- \\

                            \href{https://gcn.gsfc.nasa.gov/gcn3/24178.gcn3}{GCN24178} 
                                	&
                							2019-04-25 08:18:05 &
                                LIGO/Virgo S190425z: further analysis of INTEGRAL data &

                                    $<$2$\times$10$^{-7}$ \\

                            \href{https://gcn.gsfc.nasa.gov/gcn3/24181.gcn3}{GCN24181} 
                                 	$^\ddagger$ &
                							2019-04-25 08:18:05 &
                                LIGO/Virgo S190425z: INTEGRAL IBIS prompt observation &

                                    -- \\

                            \href{https://gcn.gsfc.nasa.gov/gcn3/24242.gcn3}{GCN24242} 
                                	&
                							2019-04-26 15:21:55 &
                                LIGO/Virgo S190426c: INTEGRAL prompt observation &

                                    $<$1.7$\times$10$^{-7}$ \\

                            \href{https://gcn.gsfc.nasa.gov/gcn3/24380.gcn3}{GCN24380} 
                                	&
                							2019-05-03 18:54:04 &
                                LIGO/Virgo S190503bf: INTEGRAL prompt observation &

                                    $<$2.4$\times$10$^{-7}$ \\

                            \href{https://gcn.gsfc.nasa.gov/gcn3/24381.gcn3}{GCN24381} 
                                	&
                							2019-05-03 17:23:08 &
                                IceCube-190503A: INTEGRAL prompt observation &

                                    $<$4$\times$10$^{-7}$ \\

                            \href{https://gcn.gsfc.nasa.gov/gcn3/24445.gcn3}{GCN24445} 
                                	&
                				                   	2019-05-10 07:07:00$^\dagger$ & 
                                   S190510g: INTEGRAL inactive at the time of the event &

                                    -- \\

                            \href{https://gcn.gsfc.nasa.gov/gcn3/24508.gcn3}{GCN24508} 
                                	&
                							2019-05-12 18:07:14 &
                                LIGO/Virgo S190512at: INTEGRAL inactive at the time of the event &

                                    -- \\

                            \href{https://gcn.gsfc.nasa.gov/gcn3/24527.gcn3}{GCN24527} 
                                	&
                							2019-05-13 20:54:28 &
                                LIGO/Virgo S190513bm: INTEGRAL prompt observation &

                                    $<$2.6$\times$10$^{-7}$ \\

                            \href{https://gcn.gsfc.nasa.gov/gcn3/24571.gcn3}{GCN24571} 
                                	&
                							2019-05-17 05:51:01 &
                                LIGO/Virgo S190517h: INTEGRAL/SPI-ACS prompt observation &

                                    $<$2.7$\times$10$^{-7}$ \\

                            \href{https://gcn.gsfc.nasa.gov/gcn3/24589.gcn3}{GCN24589} 
                                	&
                							2019-05-17 05:51:01 &
                                LIGO/Virgo S190517h: INTEGRAL prompt observation &

                                    $<$2.7$\times$10$^{-7}$ \\

                            \href{https://gcn.gsfc.nasa.gov/gcn3/24600.gcn3}{GCN24600} 
                                	&
                							2019-05-19 15:35:44 &
                                LIGO/Virgo S190519bj: INTEGRAL SPI/ACS prompt observation &

                                    $<$2.9$\times$10$^{-7}$ \\

                            \href{https://gcn.gsfc.nasa.gov/gcn3/24620.gcn3}{GCN24620} 
                                	&
                							2019-05-21 03:02:29 &
                                LIGO/Virgo S190521g: INTEGRAL SPI/ACS prompt observation &

                                    $<$2.3$\times$10$^{-7}$ \\

                            \href{https://gcn.gsfc.nasa.gov/gcn3/24624.gcn3}{GCN24624} 
                                	&
                							2019-05-21 03:02:29 &
                                LIGO/Virgo S190521g: No counterpart candidates in INTEGRAL SPI/ACS \& IBIS/PICsIT prompt observations &

                                    $<$1.7$\times$10$^{-7}$ \\

                            \href{https://gcn.gsfc.nasa.gov/gcn3/24629.gcn3}{GCN24629} 
                                	&
                							2019-05-21 07:43:59 &
                                LIGO/Virgo S190521r: No counterpart candidates in INTEGRAL SPI/ACS prompt observation &

                                    $<$4$\times$10$^{-7}$ \\

                            \href{https://gcn.gsfc.nasa.gov/gcn3/24727.gcn3}{GCN24727} 
                                 	$^\ddagger$ &
                							2019-06-02 17:59:27 &
                                LIGO/Virgo S190602aq: upper limit  in INTEGRAL SPI-ACS prompt observations &

                                    -- \\

                            \href{https://gcn.gsfc.nasa.gov/gcn3/24728.gcn3}{GCN24728} 
                                	&
                							2019-06-02 17:59:27 &
                                LIGO/Virgo S190602aq: No counterpart candidates in INTEGRAL SPI-ACS and IBIS prompt observation &

                                    $<$3.8$\times$10$^{-7}$ \\

                            \href{https://gcn.gsfc.nasa.gov/gcn3/24858.gcn3}{GCN24858} 
                                	&
                							2019-06-19 13:14:18 &
                                IceCube-190619A: No counterpart candidates in INTEGRAL SPI-ACS and IBIS prompt observation &

                                    $<$2$\times$10$^{-7}$ \\

                            \href{https://gcn.gsfc.nasa.gov/gcn3/24925.gcn3}{GCN24925} 
                                	&
                							2019-06-30 18:52:05 &
                                LIGO/Virgo S190630ag: No counterpart candidates in INTEGRAL SPI-ACS prompt observation &

                                    $<$1.9$\times$10$^{-7}$ \\

                            \href{https://gcn.gsfc.nasa.gov/gcn3/24951.gcn3}{GCN24951} 
                                	&
                							2019-07-01 20:33:06 &
                                LIGO/Virgo S190701ah: No counterpart candidates in INTEGRAL SPI-ACS prompt observation &

                                    $<$1.6$\times$10$^{-7}$ \\

                            \href{https://gcn.gsfc.nasa.gov/gcn3/24985.gcn3}{GCN24985} 
                                	&
                							2019-07-04 18:48:52 &
                                LIGO/Virgo IceCube-190704A: No counterpart candidates in INTEGRAL SPI- ACS and IBIS prompt observation &

                                    $<$2$\times$10$^{-7}$ \\

                            \href{https://gcn.gsfc.nasa.gov/gcn3/25010.gcn3}{GCN25010} 
                                	&
                							2019-07-06 22:26:41 &
                                LIGO/Virgo S190706ai: No counterpart candidates in INTEGRAL SPI-ACS and IBIS prompt observations &

                                    $<$1.7$\times$10$^{-7}$ \\

                            \href{https://gcn.gsfc.nasa.gov/gcn3/25029.gcn3}{GCN25029} 
                                	&
                							2019-07-07 09:33:26 &
                                LIGO/Virgo S190707q: No counterpart candidates in INTEGRAL SPI-ACS and IBIS prompt observations &

                                    $<$2.6$\times$10$^{-7}$ \\

                            \href{https://gcn.gsfc.nasa.gov/gcn3/25058.gcn3}{GCN25058} 
                                	&
                							2019-07-12 01:15:17 &
                                IceCube-190712A: No counterpart candidates in INTEGRAL SPI-ACS and IBIS prompt observation &

                                    $<$2$\times$10$^{-7}$ \\

                            \href{https://gcn.gsfc.nasa.gov/gcn3/25092.gcn3}{GCN25092} 
                                	&
                							2019-07-18 14:35:12 &
                                LIGO/Virgo S190718y: No counterpart candidates in INTEGRAL SPI-ACS prompt observation &

                                    $<$2.8$\times$10$^{-7}$ \\

                            \href{https://gcn.gsfc.nasa.gov/gcn3/25119.gcn3}{GCN25119} 
                                	&
                							2019-07-20 00:08:36 &
                                LIGO/Virgo S190720a: No counterpart candidates in INTEGRAL SPI-ACS and IBIS prompt observation &

                                    $<$1.8$\times$10$^{-7}$ \\

                            \href{https://gcn.gsfc.nasa.gov/gcn3/25166.gcn3}{GCN25166} 
                                	&
                							2019-07-27 06:03:33 &
                                LIGO/Virgo S190727h: No counterpart candidates in INTEGRAL SPI-ACS prompt observation &

                                    $<$2.1$\times$10$^{-7}$ \\

                            \href{https://gcn.gsfc.nasa.gov/gcn3/25171.gcn3}{GCN25171} 
                                 	$^\ddagger$ &
                							2019-07-27 06:03:33 &
                                LIGO/Virgo S190727h: possible counterpart candidate in SPI-ACS/INTEGRAL &

                                    -- \\

                            \href{https://gcn.gsfc.nasa.gov/gcn3/25189.gcn3}{GCN25189} 
                                 	$^\ddagger$ &
                							2019-07-28 06:45:10 &
                                LIGO/Virgo S190728q: SPI-ACS/INTEGRAL data analysis &

                                    -- \\

                            \href{https://gcn.gsfc.nasa.gov/gcn3/25190.gcn3}{GCN25190} 
                                	&
                							2019-07-28 06:45:10 &
                                LIGO/Virgo S190728q: No counterpart candidates in INTEGRAL SPI-ACS prompt observation &

                                    $<$2.6$\times$10$^{-7}$ \\

                            \href{https://gcn.gsfc.nasa.gov/gcn3/25232.gcn3}{GCN25232} 
                                	&
                							2019-07-30 20:50:41 &
                                IceCube 190730A: one weakly associated counterpart candidates in INTEGRAL SPI-ACS, and IBIS prompt observation &

                                    $<$1.8$\times$10$^{-7}$ \\

                            \href{https://gcn.gsfc.nasa.gov/gcn3/25309.gcn3}{GCN25309} 
                                	&
                							2019-08-06 13:20:48 &
                                HAWC-190806A: upper limits from INTEGRAL SPI-ACS and IBIS prompt observation &

                                    $<$2$\times$10$^{-7}$ \\

                            \href{https://gcn.gsfc.nasa.gov/gcn3/25323.gcn3}{GCN25323} 
                                	&
                							2019-08-14 21:10:39 &
                                LIGO/Virgo S190814bv: No counterpart candidates in INTEGRAL SPI-ACS prompt observation &

                                    $<$3.1$\times$10$^{-7}$ \\

                            \href{https://gcn.gsfc.nasa.gov/gcn3/25403.gcn3}{GCN25403} 
                                	&
                							2019-08-19 17:34:24 &
                                IceCube-190819A: No counterpart candidates in INTEGRAL SPI-ACS prompt observation &

                                    $<$3.5$\times$10$^{-7}$ \\

                            \href{https://gcn.gsfc.nasa.gov/gcn3/25407.gcn3}{GCN25407} 
                                	&
                				                   	2019-08-20 09:08:59$^\dagger$ & 
                                   Fermi GBM-190816: Location-dependent upper limits from INTEGRAL/SPI-ACS prompt observation &

                                    $<$1.8$\times$10$^{-7}$ \\

                            \href{https://gcn.gsfc.nasa.gov/gcn3/25500.gcn3}{GCN25500} 
                                	&
                							2019-08-28 06:34:05 &
                                LIGO/Virgo S190828j: No counterpart candidates in INTEGRAL SPI-ACS prompt observation &

                                    $<$2$\times$10$^{-7}$ \\

                            \href{https://gcn.gsfc.nasa.gov/gcn3/25505.gcn3}{GCN25505} 
                                	&
                							2019-08-28 06:55:09 &
                                LIGO/Virgo S190828l: No counterpart candidates in INTEGRAL SPI-ACS prompt observation &

                                    $<$1.8$\times$10$^{-7}$ \\

                            \href{https://gcn.gsfc.nasa.gov/gcn3/25605.gcn3}{GCN25605} 
                                	&
                							2019-09-01 23:31:01 &
                                LIGO/Virgo S190901ap: No counterpart candidates in INTEGRAL SPI-ACS,prompt observation &

                                    $<$1.7$\times$10$^{-7}$ \\

                            \href{https://gcn.gsfc.nasa.gov/gcn3/25698.gcn3}{GCN25698} 
                                	&
                							2019-09-10 01:26:19 &
                                LIGO/Virgo S190910d: No counterpart candidates in INTEGRAL SPI-ACS prompt observation &

                                    $<$2.2$\times$10$^{-7}$ \\

                            \href{https://gcn.gsfc.nasa.gov/gcn3/25709.gcn3}{GCN25709} 
                                	&
                							2019-09-10 08:29:58 &
                                LIGO/Virgo S190910h: No counterpart candidates in INTEGRAL SPI-ACS prompt observation &

                                    $<$3.1$\times$10$^{-7}$ \\

                            \href{https://gcn.gsfc.nasa.gov/gcn3/25755.gcn3}{GCN25755} 
                                	&
                							2019-09-15 23:57:02 &
                                LIGO/Virgo S190915ak: No counterpart candidates in INTEGRAL SPI-ACS prompt observation &

                                    $<$2$\times$10$^{-7}$ \\

                            \href{https://gcn.gsfc.nasa.gov/gcn3/25774.gcn3}{GCN25774} 
                                	&
                							2019-09-17 01:14:19 &
                                HAWC-190916A: No counterpart candidates in INTEGRAL SPI-ACS prompt observation &

                                    $<$3.3$\times$10$^{-7}$ \\

                            \href{https://gcn.gsfc.nasa.gov/gcn3/25803.gcn3}{GCN25803} 
                                	&
                							2019-09-22 09:42:45 &
                                IceCube-190922A: No counterpart candidates in INTEGRAL SPI-ACS prompt observation &

                                    $<$3.8$\times$10$^{-7}$ \\

                            \href{https://gcn.gsfc.nasa.gov/gcn3/25809.gcn3}{GCN25809} 
                                	&
                				                   	2019-09-23 13:11:41$^\dagger$ & 
                                   IceCube-190922B: INTEGRAL was inactive at the time of the event &

                                    -- \\

                            \href{https://gcn.gsfc.nasa.gov/gcn3/25815.gcn3}{GCN25815} 
                                	&
                							2019-09-23 12:55:59 &
                                LIGO/Virgo S190923y: No counterpart candidates in INTEGRAL SPI-ACS prompt observation &

                                    $<$3.4$\times$10$^{-7}$ \\

                            \href{https://gcn.gsfc.nasa.gov/gcn3/25825.gcn3}{GCN25825} 
                                	&
                							2019-09-23 12:55:59 &
                                LIGO/Virgo S190923y: INTEGRAL IBIS coded-masked imaging observations of part of the localization area and further all-sky upper limits &

                                    $<$3.4$\times$10$^{-7}$ \\

                            \href{https://gcn.gsfc.nasa.gov/gcn3/25837.gcn3}{GCN25837} 
                                	&
                							2019-09-24 02:18:46 &
                                LIGO/Virgo S190924h: No counterpart candidates in INTEGRAL SPI-ACS and IBIS prompt observation &

                                    $<$4.2$\times$10$^{-7}$ \\

                            \href{https://gcn.gsfc.nasa.gov/gcn3/25872.gcn3}{GCN25872} 
                                	&
                							2019-09-30 13:35:41 &
                                LIGO/Virgo S190930s: No counterpart candidates in INTEGRAL SPI-ACS prompt observation &

                                    $<$1.8$\times$10$^{-7}$ \\

                            \href{https://gcn.gsfc.nasa.gov/gcn3/25880.gcn3}{GCN25880} 
                                	&
                							2019-09-30 14:34:07 &
                                LIGO/Virgo S190930t: No counterpart candidates in INTEGRAL SPI-ACS prompt observation &

                                    $<$3.3$\times$10$^{-7}$ \\

                            \href{https://gcn.gsfc.nasa.gov/gcn3/25914.gcn3}{GCN25914} 
                                	&
                							2019-10-01 20:09:18 &
                                IceCube-191001A: No counterpart candidates in INTEGRAL SPI-ACS prompt observation &

                                    $<$2.9$\times$10$^{-7}$ \\

                            \href{https://gcn.gsfc.nasa.gov/gcn3/26187.gcn3}{GCN26187} 
                                	&
                							2019-11-05 14:35:21 &
                                LIGO/Virgo S191105e: No counterpart candidates in INTEGRAL SPI-ACS and,IBIS prompt observation &

                                    $<$2.3$\times$10$^{-7}$ \\

                            \href{https://gcn.gsfc.nasa.gov/gcn3/26207.gcn3}{GCN26207} 
                                	&
                							2019-11-09 01:07:17 &
                                LIGO/Virgo S191109d: No counterpart candidates in INTEGRAL SPI-ACS,prompt observation &

                                    $<$5.3$\times$10$^{-7}$ \\

                            \href{https://gcn.gsfc.nasa.gov/gcn3/26225.gcn3}{GCN26225} 
                                	&
                							2019-11-10 23:06:44 &
                                LIGO/Virgo S191110af: No counterpart candidates in INTEGRAL SPI-ACS, prompt observation &

                                    $<$2.8$\times$10$^{-7}$ \\

                            \href{https://gcn.gsfc.nasa.gov/gcn3/26232.gcn3}{GCN26232} 
                                	&
                							2019-11-10 23:06:44 &
                                LIGO/Virgo S191110af: No counterpart candidates in INTEGRAL SPI-ACS, and IBIS prompt observation &

                                    $<$5$\times$10$^{-7}$ \\

                            \href{https://gcn.gsfc.nasa.gov/gcn3/26259.gcn3}{GCN26259} 
                                	&
                							2019-11-19 01:01:29 &
                                IceCube-191119A - No counterpart candidates in INTEGRAL SPI-ACS prompt observation &

                                    $<$4.9$\times$10$^{-7}$ \\

                            \href{https://gcn.gsfc.nasa.gov/gcn3/26281.gcn3}{GCN26281} 
                                	&
                				                   	2019-11-23 14:32:42$^\dagger$ & 
                                   IceCube-191122A: INTEGRAL was inactive at the time of the event &

                                    -- \\

                            \href{https://gcn.gsfc.nasa.gov/gcn3/26311.gcn3}{GCN26311} 
                                	&
                							2019-11-29 13:40:29 &
                                LIGO/Virgo S191129u: no INTEGRAL data available &

                                    -- \\

                            \href{https://gcn.gsfc.nasa.gov/gcn3/26333.gcn3}{GCN26333} 
                                	&
                							2019-12-04 17:15:26 &
                                LIGO/Virgo S191204r: No counterpart candidates in INTEGRAL SPI-ACS prompt observation &

                                    $<$3.4$\times$10$^{-7}$ \\

                            \href{https://gcn.gsfc.nasa.gov/gcn3/26339.gcn3}{GCN26339} 
                                	&
                							2019-12-04 22:46:11 &
                                IceCube-191204A: No counterpart candidates in INTEGRAL SPI-ACS prompt observation &

                                    $<$2.6$\times$10$^{-7}$ \\

                            \href{https://gcn.gsfc.nasa.gov/gcn3/26351.gcn3}{GCN26351} 
                                	&
                							2019-12-05 21:52:08 &
                                LIGO/Virgo S191205ah: No counterpart candidates in INTEGRAL SPI-ACS prompt observation &

                                    $<$1.8$\times$10$^{-7}$ \\

                            \href{https://gcn.gsfc.nasa.gov/gcn3/26401.gcn3}{GCN26401} 
                                	&
                							2019-12-13 04:34:08 &
                                LIGO/Virgo S191213g: No counterpart candidates in INTEGRAL SPI-ACS prompt observation &

                                    $<$5.2$\times$10$^{-7}$ \\

                            \href{https://gcn.gsfc.nasa.gov/gcn3/26436.gcn3}{GCN26436} 
                                	&
                							2019-12-15 11:09:57 &
                                IceCube-191215A: No counterpart candidates in INTEGRAL SPI-ACS prompt observation &

                                    $<$3.7$\times$10$^{-7}$ \\

                            \href{https://gcn.gsfc.nasa.gov/gcn3/26442.gcn3}{GCN26442} 
                                	&
                							2019-12-15 22:30:52 &
                                LIGO/Virgo S191215w: No counterpart candidates in INTEGRAL SPI-ACS prompt observation &

                                    $<$3.3$\times$10$^{-7}$ \\

                            \href{https://gcn.gsfc.nasa.gov/gcn3/26456.gcn3}{GCN26456} 
                                	&
                							2019-12-16 21:33:38 &
                                LIGO/Virgo S191216ap: no INTEGRAL observation &

                                    -- \\

                            \href{https://gcn.gsfc.nasa.gov/gcn3/26545.gcn3}{GCN26545} 
                                	&
                							2019-12-22 03:35:37 &
                                LIGO/Virgo S191222n: INTEGRAL was inactive at the time of the event &

                                    -- \\

                            \href{https://gcn.gsfc.nasa.gov/gcn3/26621.gcn3}{GCN26621} 
                                	&
                							2019-12-31 11:00:06 &
                                IceCube-191231A: No counterpart candidates in INTEGRAL SPI-ACS and IBIS prompt observation &

                                    $<$2.1$\times$10$^{-7}$ \\

                            \href{https://gcn.gsfc.nasa.gov/gcn3/26644.gcn3}{GCN26644} 
                                	&
                							2020-01-05 16:24:26 &
                                LIGO/Virgo S200105ae: No counterpart candidates in INTEGRAL SPI-ACS prompt observation &

                                    $<$2.2$\times$10$^{-7}$ \\

                            \href{https://gcn.gsfc.nasa.gov/gcn3/26666.gcn3}{GCN26666} 
                                	&
                							2020-01-07 09:42:18 &
                                IceCube-200107A: No counterpart candidates in INTEGRAL SPI-ACS prompt observation &

                                    $<$2.8$\times$10$^{-7}$ \\

                            \href{https://gcn.gsfc.nasa.gov/gcn3/26683.gcn3}{GCN26683} 
                                	&
                							2020-01-08 09:30:14 &
                                Fermi-LAT ANTARES coincidence: No counterpart candidates in INTEGRAL SPI-ACS and IBIS prompt observation &

                                    $<$2$\times$10$^{-7}$ \\

                            \href{https://gcn.gsfc.nasa.gov/gcn3/26698.gcn3}{GCN26698} 
                                	&
                				                   	2020-01-10 08:48:10$^\dagger$ & 
                                   IceCube-200109A: INTEGRAL was inactive at the time of the event &

                                    -- \\

                            \href{https://gcn.gsfc.nasa.gov/gcn3/26721.gcn3}{GCN26721} 
                                	&
                							2020-01-12 15:58:38 &
                                LIGO/Virgo S200112r: INTEGRAL was inactive at the time of the event &

                                    -- \\

                            \href{https://gcn.gsfc.nasa.gov/gcn3/26743.gcn3}{GCN26743} 
                                	&
                							2020-01-14 02:08:18 &
                                LIGO/Virgo S200114f: No counterpart candidates in INTEGRAL SPI-ACS and IBIS prompt observation &

                                    $<$3.1$\times$10$^{-7}$ \\

                            \href{https://gcn.gsfc.nasa.gov/gcn3/26766.gcn3}{GCN26766} 
                                	&
                							2020-01-15 04:23:09 &
                                LIGO/Virgo S200115j: INTEGRAL was inactive at the time of the event &

                                    -- \\

                            \href{https://gcn.gsfc.nasa.gov/gcn3/26804.gcn3}{GCN26804} 
                                	&
                							2020-01-17 11:08:29 &
                                IceCube-200117A: no counterpart candidates in INTEGRAL SPI-ACS and,IBIS prompt observation &

                                    $<$3.3$\times$10$^{-7}$ \\

                            \href{https://gcn.gsfc.nasa.gov/gcn3/26838.gcn3}{GCN26838} 
                                	&
                							2020-01-20 18:48:18 &
                                IceCube-200120A: no counterpart candidates in INTEGRAL SPI-ACS and,IBIS prompt observation &

                                    $<$2$\times$10$^{-7}$ \\

                            \href{https://gcn.gsfc.nasa.gov/gcn3/26908.gcn3}{GCN26908} 
                                	&
                							2020-01-28 02:20:11 &
                                LIGO/Virgo S200128d: No counterpart candidates in INTEGRAL SPI-ACS prompt observation &

                                    $<$2.6$\times$10$^{-7}$ \\

                            \href{https://gcn.gsfc.nasa.gov/gcn3/26928.gcn3}{GCN26928} 
                                	&
                							2020-01-29 06:54:58 &
                                LIGO/Virgo S200129m: No counterpart candidates in INTEGRAL SPI-ACS prompt observation &

                                    $<$4.5$\times$10$^{-7}$ \\

                            \href{https://gcn.gsfc.nasa.gov/gcn3/27019.gcn3}{GCN27019} 
                                	&
                							2020-02-08 13:01:17 &
                                LIGO/Virgo S200208q: No counterpart candidates in INTEGRAL SPI-ACS prompt observation &

                                    $<$2.3$\times$10$^{-7}$ \\

                            \href{https://gcn.gsfc.nasa.gov/gcn3/27050.gcn3}{GCN27050} 
                                	&
                							2020-02-13 04:10:40 &
                                LIGO/Virgo S200213t: No counterpart candidates in INTEGRAL SPI-ACS prompt observation &

                                    $<$4.8$\times$10$^{-7}$ \\

                            \href{https://gcn.gsfc.nasa.gov/gcn3/27131.gcn3}{GCN27131} 
                                	&
                							2020-02-19 09:44:15 &
                                LIGO/Virgo S200219ac: No counterpart candidates in INTEGRAL SPI-ACS and IBIS prompt observation &

                                    $<$2.4$\times$10$^{-7}$ \\

                            \href{https://gcn.gsfc.nasa.gov/gcn3/27190.gcn3}{GCN27190} 
                                	&
                							2020-02-24 22:22:34 &
                                LIGO/Virgo S200224ca: No counterpart candidates in INTEGRAL SPI-ACS prompt observation &

                                    $<$3$\times$10$^{-7}$ \\

                            \href{https://gcn.gsfc.nasa.gov/gcn3/27198.gcn3}{GCN27198} 
                                	&
                							2020-02-25 06:04:21 &
                                LIGO/Virgo S200225q: No counterpart candidates in INTEGRAL SPI-ACS and IBIS prompt observation &

                                    $<$2.1$\times$10$^{-7}$ \\

                            \href{https://gcn.gsfc.nasa.gov/gcn3/27237.gcn3}{GCN27237} 
                                	&
                							2020-02-27 05:36:31 &
                                IceCube-200227A: No counterpart candidates in INTEGRAL SPI-ACS and IBIS prompt observation &

                                    $<$2.9$\times$10$^{-7}$ \\

                            \href{https://gcn.gsfc.nasa.gov/gcn3/27285.gcn3}{GCN27285} 
                                	&
                							2020-03-02 01:58:11 &
                                LIGO/Virgo S200302c: No counterpart candidates in INTEGRAL SPI-ACS and IBIS prompt observation &

                                    $<$3.2$\times$10$^{-7}$ \\

                            \href{https://gcn.gsfc.nasa.gov/gcn3/27360.gcn3}{GCN27360} 
                                	&
                							2020-03-11 11:58:53 &
                                LIGO/Virgo S200311bg: No counterpart candidates in INTEGRAL SPI-ACS,prompt observation &

                                    $<$4.5$\times$10$^{-7}$ \\

                            \href{https://gcn.gsfc.nasa.gov/gcn3/27397.gcn3}{GCN27397} 
                                	&
                							2020-03-16 21:57:56 &
                                LIGO/Virgo S200316bj: No counterpart candidates in INTEGRAL SPI-ACS, and IBIS prompt observation &

                                    $<$4.3$\times$10$^{-7}$ \\

                            \href{https://gcn.gsfc.nasa.gov/gcn3/27537.gcn3}{GCN27537} 
                                	&
                							2020-04-10 23:19:55 &
                                IceCube-200410A: Possible associated excess in INTEGRAL SPI-ACS prompt observation &

                                    $<$5.2$\times$10$^{-7}$ \\

                            \href{https://gcn.gsfc.nasa.gov/gcn3/27613.gcn3}{GCN27613} 
                                	&
                							2020-04-21 00:35:24 &
                                IceCube-200421A: No counterpart candidates in INTEGRAL SPI-ACS and IBIS prompt observation &

                                    $<$4.9$\times$10$^{-7}$ \\

                            \href{https://gcn.gsfc.nasa.gov/gcn3/27652.gcn3}{GCN27652} 
                                	&
                							2020-04-25 23:26:46 &
                                IceCube-200425A: No counterpart candidates in INTEGRAL SPI-ACS prompt observation &

                                    $<$2.8$\times$10$^{-7}$ \\

                            \href{https://gcn.gsfc.nasa.gov/gcn3/27668.gcn3}{GCN27668} 
                                	&
                				                   	2020-04-29 09:30:38$^\dagger$ & 
                                   SGR 1935+2154: INTEGRAL hard X-ray counterpart of radio burst &

                                    -- \\

                            \href{https://gcn.gsfc.nasa.gov/gcn3/27720.gcn3}{GCN27720} 
                                	&
                							2020-05-12 07:31:27 &
                                IceCube-200512A: No counterpart candidates in INTEGRAL SPI-ACS prompt observation &

                                    $<$4.9$\times$10$^{-7}$ \\

                            \href{https://gcn.gsfc.nasa.gov/gcn3/27866.gcn3}{GCN27866} 
                                	&
                							2020-05-30 07:54:29 &
                                IceCube-200530A: No counterpart candidates in INTEGRAL SPI-ACS prompt observation &

                                    $<$2.7$\times$10$^{-7}$ \\

                            \href{https://gcn.gsfc.nasa.gov/gcn3/27952.gcn3}{GCN27952} 
                                	&
                							2020-06-14 12:41:21 &
                                IceCube-200614A: No counterpart candidates in INTEGRAL SPI-ACS prompt observation &

                                    $<$2.7$\times$10$^{-7}$ \\

                            \href{https://gcn.gsfc.nasa.gov/gcn3/27954.gcn3}{GCN27954} 
                                	&
                							2020-06-15 14:49:17 &
                                IceCube-200615A: No counterpart candidates in INTEGRAL SPI-ACS prompt observation &

                                    $<$3.5$\times$10$^{-7}$ \\

                            \href{https://gcn.gsfc.nasa.gov/gcn3/28009.gcn3}{GCN28009} 
                                	&
                				                   	2020-06-22 14:04:20$^\dagger$ & 
                                   IceCube-200620A: INTEGRAL was inactive at the time of the event &

                                    -- \\

	\end{longtable}
}